%% file: paper.tex
\documentclass[fleqn,usenatbib]{mnras}

\usepackage{newtxtext,newtxmath}

\usepackage[T1]{fontenc}
\usepackage{ae,aecompl}

\usepackage{pdflscape}
\usepackage{adjustbox}

\usepackage{breqn}
\usepackage{catchfile}

\usepackage{graphicx}	
\usepackage{amsmath}	
\usepackage{siunitx}

\title[A sample of 318 new young stars]{A spectroscopically confirmed Gaia-selected sample of 318 new young stars within $\sim$200~pc}

\author[{\v Z}erjal et al.]{
Maru{\v s}a {\v Z}erjal,$^{1}$\thanks{E-mail: marusa.zerjal@anu.edu.au}
Adam D. Rains,$^{1}$
Michael J. Ireland,$^{1}$
George Zhou,$^{2,3}$
\newauthor 
Jens Kammerer$^{1,4}$, 
Alex Wallace$^{1}$, 
Brendan Orenstein$^{1}$, 
Thomas Nordlander$^{1,5}$,
\newauthor 
Harrison Abbot$^{1}$,
Seo-Won Chang$^{1,6,7,8}$
\\
$^{1}$Research School of Astronomy \& Astrophysics, Australian National University, ACT 2611, Australia\\
$^{2}$Center for Astrophysics, Harvard \& Smithsonian, 60 Garden St., Cambridge, MA 02138, USA\\
$^{3}$Hubble fellow \\
$^{4}$European Southern Observatory, Karl-Schwarzschild-Str 2, 85748, Garching, Germany\\
$^5$ ARC Centre of Excellence for All Sky Astrophysics in 3 Dimensions (ASTRO 3D)\\
$^{6}$ARC Centre of Excellence for Gravitational Wave Discovery (OzGrav), Australia\\
$^{7}$SNU Astronomy Research Center, Seoul National University, 1 Gwanak-rho, Gwanak-gu, Seoul 08826, Korea\\
$^{8}$Astronomy program, Dept. of Physics \& Astronomy, Seoul National University, 1 Gwanak-rho, Gwanak-gu, Seoul 08826, Korea\\
}

\begin{document}

\maketitle

\begin{abstract}
In the Gaia era, the majority of stars in the Solar neighbourhood have parallaxes and proper motions precisely determined while spectroscopic age indicators are still missing for a large fraction of low-mass young stars. In this work we select 756 overluminous late K and early M young star candidates in the southern sky and observe them over 64 nights with the ANU 2.3m Telescope at Siding Spring Observatory using the Echelle (R=24,000) and Wide Field spectrographs  (WiFeS, R=3000-7000). Our selection is kinematically unbiased to minimize the preference against low-mass members of stellar associations that dissipate first, and to include potential members of diffuse components. We provide measurements of H$\alpha$ and calcium~H\&K emission, as well as lithium absorption line, that enable identification of stars as young as $\sim$10-30~Myr which is a typical age of a stellar association.  We report on 346 stars showing a detectable lithium line, 318 of which are not found in the known catalogs of young stars. We also report 126 additional stars in our sample which have no detectable lithium but signs of stellar activity indicating youth.
Radial velocities are determined for WiFeS spectra with a precision of 3.2~$\mathrm{km\;s^{-1}}$ and 1.5~$\mathrm{km\;s^{-1}}$ for the Echelle sample.



\end{abstract}




\section{Introduction}




Star-forming regions in the Galaxy are distributed in a complex web of filaments that resemble a highly hierarchical network \cite[e.g.][]{2018A&A...610A..77H, 2010A&A...518L.100M, 2010A&A...518L.102A}. While open clusters are typically found in the densest parts of the structure, nearly 90\% of newborn stars become gravitationally unbound soon after the birth due to their dynamic interactions. Such loose ensembles of dispersing coeval stars are observed as stellar associations that keep the kinematic imprint of their local birth site up to $\sim$30~Myr before they become a part of the Galactic disk \citep{2019ARA&A..57..227K}. Because such groups of hundreds to thousands of stellar siblings were born from the same molecular cloud, they all have similar surface abundances \citep{2007AJ....133..694D}. These moving groups are thus the fossil records of the Galaxy that have a potential to link together star formation sites with the larger structures of the disk. They resemble an ideal laboratory to study a wide variety of important topics, from star- and planetary formation environments, the initial mass function and sequentially triggered star formation to dynamical processes that lead to the evaporation and finally the dispersal of an association.

A reliable reconstruction of stellar associations is thus of critical importance. While observations from the Hipparcos space astrometry mission allowed a major improvement in the search of overdensities in the kinematic phase space using stellar positions, parallaxes and proper motions \citep{1999AJ....117..354D}, it is high precision measurements from the Gaia space telescope -- including radial velocities for a subset of 7,000,000 stars -- that is revolutionising Galactic astrophysics \citep{2018A&A...616A...1G}. It has facilitated numerous attempts to study young stars above the main sequence and identify new members of the known moving groups in the Solar neighbourhood 
\cite[e.g.][]{2018ApJ...862..138G, 2020MNRAS.491..215B}. Additionally
\citet{2018A&A...618A..93C} studied young populations on much larger Galactic scales and reported on the discovery of $\sim$1500 clusters. 

Although a selection of the candidate members of a particular moving group is often based on the cuts in the kinematic space \cite[e.g.][]{2020arXiv200204801U}, 
the true nature of these groups appear to be diffuse due to their gradual dispersal. \citet{2019A&A...621L...3M} recently described extended structures emerging as the tidal streams of the nearby Hyades cluster, while \citet{2019A&A...623A.112D} found 11,000 pre-main sequence members of the Scorpius-Centaurus OB2 association residing in both compact and diffuse populations. Kinematic cuts in such cases are prone to be biased against the low-mass stars that are most likely to evaporate first.

Numerous works on young associations rely on multi-dimensional clustering algorithms. For example, \cite[e.g.][]{2019AJ....158..122K} report on the discovery of 1,900 clusters and comoving groups within 1~kpc with HDBSCAN (Hierarchival Density-Based Spatial Clustering of Applications with Noise described by \citealp{campello_density-based_2013}). However, the arrival of the Gaia's high precision parallaxes and proper motions enables reliable orbital simulations for the first time. For instance, \citet{2019MNRAS.489.3625C} were able to model an association at its birth time using Chronostar, perform its orbital trace-forward and blindly reconstruct the known Beta Pictoris association, reliably determine its members and, importantly, its kinematic age.

Stellar age is, besides the kinematics, one of the decisive parameters in the characterization of the young moving groups. Parallaxes of nearby stars with uncertainties better than 10\% enable the placement of stellar populations on the color-magnitude diagram. However, due to the numerous effects including the evolutionary model uncertainties and inflated radii on low-mass end of the population, the presence of binaries, background contamination and spread due to metallicity effects, and the variability of young stars, isochronal dating techniques remain non-trivial. 

While gyrochronology relies on the multiple photometric measurements to determine the rotation period of a star, it is spectroscopic youth indicators that require only one observation for the estimation of stellar age. 
Spectroscopic features of solar-like and cooler young stars up to the solar age are straightforward to observe. They emerge from the processes related to the magnetic activity of a star and manifest themselves in the excess emission in calcium H\&K lines (Ca~II~H\&K, 3969 and 3934~\AA; \citealt{2008ApJ...687.1264M}), H$\alpha$ line (6563~\AA; \citealt{2005A&A...431..329L}) and infrared calcium triplet (Ca~II~IRT; 8498, 8542 and 8662~\AA; \citealt{2013ApJ...776..127Z}). \citealp{2008ApJ...687.1264M} describe an age--activity relation that estimates age from the Ca~II~H\&K emission in the range from $\sim$10~Myr up to 10~Gyr, although \citealp{2013A&A...551L...8P} has shown later that there is no decay in chromospheric activity beyond 2~Gyr. The decline of the emission rate is the fastest in the youngest stars. Despite the variable nature of magnetic activity, especially in the pre-main sequence stars, it is easy to differentiate between stars of a few 10 and a few 100~Myr.
On the other hand, the presence of the lithium 6708~\AA~ line in GKM dwarfs directly indicates their youth and is a good age estimator for stars between 10-30~Myr -- which is a typical age of a stellar association.

Follow-up observations with the goal to detect the lithium line in young candidates have been performed by \citet{2019ApJ...877...60B} (who found lithium in 58 stars) 
while \citet{2009A&A...508..833D} 
report on the lithium measurements for $\sim$400 stars.
Over 3000 young K and M stars with a detectable lithium 6708~\AA~ line have recently been identified in the GALAH dataset  \citep{2019MNRAS.484.4591Z}. While the majority of young early K dwarfs in the GALAH sample have practically settled on the main sequence, young late K and M stars with a detectable lithium line still reside 1~magnitude or more above the main sequence. 
\citealp{2015MNRAS.448.2737R} have kinematically and photometrically selected candidate members of the Upper-Scorpius association and discovered 237 new members by the presence of lithium absorption.



In the Gaia era, the majority of stars in the Solar neighbourhood have parallaxes and proper motions precisely determined while spectroscopic age indicators and precise radial velocities are missing for a large fraction of low-mass young stars. Large spectroscopic surveys, such as GALAH \citep{2020arXiv201102505B}, typically avoid the crowded Galactic plane where most of the young stars reside.
This work aims to fill the gap and presents spectroscopic observations, their age indicators and radial velocities of 799 young star candidates within 200~pc with no pre-existing lithium measurements. 
Section \ref{sec.data} describes the kinematically unbiased selection of all overluminous late K and early M stars within 200~pc. 
We measure equivalent widths of the lithium absorption lines and the excess flux in Ca~II~H\&K and H$\alpha$ lines, as described in Section \ref{sec.youth_indicators}. Section \ref{sec.discussion} discusses age estimation and strategy success. The dataset is accompanied with radial velocities. Concluding remarks are given in Section \ref{sec.conclusions}.


\section{Data} \label{sec.data}
\subsection{Selection function}
Candidate young stars with Gaia magnitudes $10<G<14.5$ were selected from the \textit{Gaia}~DR2 catalogue \citep{2018A&A...616A...1G}. We focused only on the low-mass end of the distribution. The selection was based on their overluminosities in the colour-magnitude diagram. The colour index was chosen to be BP-W1 because it gives the narrowest main sequence with overluminous stars clearly standing out. BP is taken from \citet{2018A&A...616A...1G} and is described in more detail by \citet{2018A&A...616A...4E} while W1 is from \citet{2014yCat.2328....0C}. The relation used as a lower main sequence parametrisation $G(c)$
\begin{dmath}
G(c) = 4.717 \times  10^{-3} \; c^5 -0.149 \; c^4 + 1.662 \; c^3 - 8.374 \; c^2 + 20.728 \; c - 14.129
\end{dmath}
where G is absolute Gaia G magnitude and $c$=BP-W1 is described in more detail in \citet{2019MNRAS.484.4591Z} together with the arguments leading to the choice of BP-W1 being the best colour index for this purpose. 
The colour-- temperature relation is determined from synthetic spectra while the temperature-spectral type relation is based on \citet{2013ApJS..208....9P}\footnote{In the version from 2018.08.02, available online: \url{http://www.pas.rochester.edu/~emamajek/EEM_dwarf_UBVIJHK_colors_Teff.txt}}.

Our criteria further exclude older stars and keep only objects that are found 1~magnitude or more above the main sequence. This approach largely avoids main sequence binaries (at most 0.75~mag above the main sequence). 
The sample was color cut to include only stars between 3$<$BP-W1$<$5.6. This limit corresponds to K5-M3 dwarfs with $T_{\rm eff}=3400$--4400\,K and allows the optimisation of the observation strategy and a focus on the cool pre-main sequence objects with the fastest lithium depletion rate. The blue limit is chosen so that it minimises the contamination with subgiants but keeps most of the late K dwarfs in the sample. The red limit is set on the steep region of the lithium isochrones that divides early M dwarfs with the fast depletion processes from those cooler ones that need more than 100~Myr to show a significant change in lithium.
The upper luminosity boundary 
\begin{equation}
G > G(c) - (1.33 c -3)
\end{equation}
rejects giants from the sample.

Since all stars disperse with time in the kinematic parameter space, young objects are found only in regions with low velocities. To avoid the kinematic bias towards the pre-selected clumps of young stars in the velocity parameter space that disfavors the low-mass stars, and to remove old stars, we compute the mean $UVW$ value of the sample and keep all objects within ($\pm$15, $\pm$15, $\pm$10)~$\mathrm{km\,s^{-1}}$ of the median $UVW$ = (-11.90, 215.77, 0.19)~$\mathrm{km\,s^{-1}}$. 
No kinematic cut was performed on stars that have no radial velocities available in the Gaia catalogue \citep{2018A&A...616A...6S}. 

\begin{figure}
\includegraphics[width=\columnwidth]{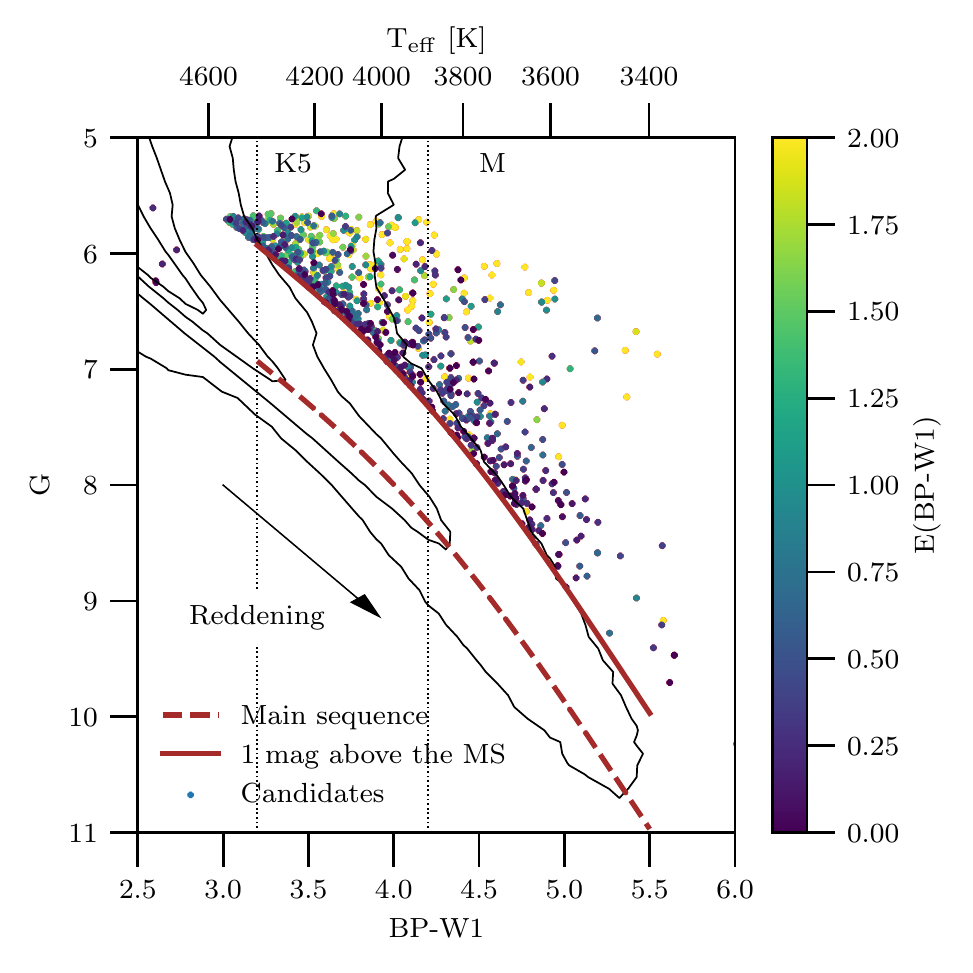}
\caption{Colour-magnitude diagram with candidate young stars and their reddening estimated in this work. 
Details on the reddening estimation are described in Sec. \ref{sec.reddening}. The most crowded region ($\sim$ K5 dwarfs) is contaminated with reddened hotter stars while M dwarfs show less contamination due to their proximity.
Red lines denote the main sequence (dashed line) and the selection function 1 magnitude above (solid line).
Contours show the density of stars in the Gaia catalog.
}
\label{fig.cmd}
\end{figure}

A declination cut with $\delta<30\,\mathrm{deg}$ eliminated objects not visible from the Siding Spring Observatory, Australia, where the observations took place.
Known young stars 
from the Simbad database and stars observed with the GALAH survey \citep{2018MNRAS.478.4513B} were removed from the list to maximise survey efficiency at detecting new young stars. This selection results in 799 candidate stars. 
Finally, our sample of stars described in this work includes observations of 756 candidate objects from this list. 
A color-magnitude diagram with all the candidates is shown in Figure \ref{fig.cmd}. Parallaxes are taken from Gaia~DR2 \citep{2018A&A...616A...1G}. 


\subsection{Observations}
\begin{figure*}
\includegraphics[width=\linewidth]{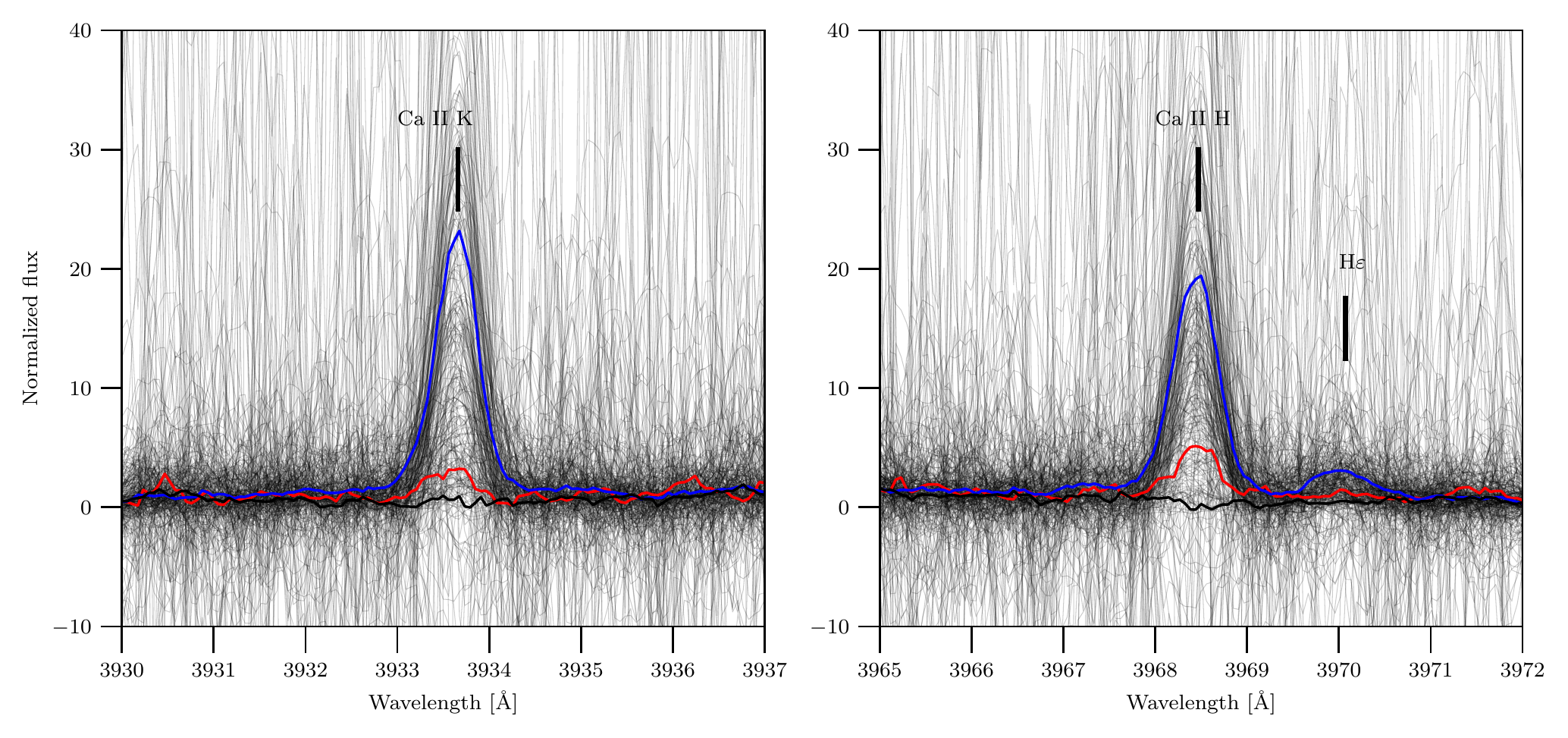}
\caption{Calcium lines in the Echelle spectra. Strong emission lines are detectable despite a low signal-to-noise ratio. There is an indication of a weak Balmer emission line at 3970~\AA. The red line is an average spectrum with a marginally detectable calcium emission while the blue line represents an average very active spectrum. Thick black line is a median inactive spectrum. Spectra in this plot were convolved with a smoothing kernel with the of width 7 for noise reduction purposes.}
\label{fig.echelle_calcium}
\end{figure*}

Observations were carried out between November 2018 and October 2019 over 64 nights with the ANU 2.3m telescope at Siding Spring Observatory. 
In order to achieve better radial velocity precision, 349 stars brighter than G=12.5 were observed with the slit-fed Echelle spectrograph in the Nasmyth focus, covering wavelengths between $\sim$3900 and $\sim$6750~\AA~ at R=24,000. Exposure times were between 600~sec for the brightest and 1800~sec for the faintest objects, resulting in a typical S/N of 20 in the order containing the H$\alpha$ line. Blue wavelengths with the calcium H\&K lines have poor S/N but clearly show strong emission above the continuum when present (Fig. \ref{fig.echelle_calcium}). The spectra were reduced as per \citet{2014MNRAS.437.2831Z}. Wavelength calibration was provided by bracketing Thorium-Argon lamp exposures.

Fainter stars (449) between $12.5<G<14.5$ were observed with the Wide Field Spectrograph (WiFeS; \citealt{2007Ap&SS.310..255D}), namely with resolving power of 3000 in the blue and 7000 in the red, covering 3500-7000~\AA. 
We typically used a RT480 beam splitter.
Typical exposure times were 5~minutes per star that resulted in the median S/N of 13 and 31 for the blue and the red band, respectively. Thorium-Argon lamp frames were taken every hour to enable wavelength calibration. WiFeS spectra were reduced with a standard PyWiFeS package \citep{2014Ap&SS.349..617C}, updated to be better suited for stellar reductions of a large number of nights.

\subsection{Synthetic Spectra} \label{sec.synthetic}
For computation of radial velocities and parameter estimation, we use a template grid of 1D LTE spectra that was previously described by \citet{2019MNRAS.488L.109N}. 
Briefly, spectra were computed using the TURBOSPECTRUM code (v15.1; \citealt{1998A&A...330.1109A, 2012ascl.soft05004P}) and MARCS model atmospheres \citep{2008A&A...486..951G}. 
For models with $\log\,g > 3.5$, we use $v_{\rm mic} = 1\,{\rm km\,s^{-1}}$; for models with $\log\,g \le 3.5$, we use $v_{\rm mic} = 2\,{\rm km\,s^{-1}}$ and perform the radiative transfer calculations under spherical symmetry taking into account continuum scattering.
The spectra are computed with a sampling step of $1\,$km$\,$s$^{-1}$, corresponding to a resolving power $R\sim300\,000$. We adopt the solar chemical composition and isotopic ratios from \citet{2009ARA&A..47..481A}, except for an alpha enhancement that varies linearly from $\text[\alpha / \text{Fe}] = 0$ when $\rm [Fe/H] \ge 0$ to $\text[\alpha/\text{Fe}] = +0.4$ when $\rm [Fe/H] \le -1$. 
We use a selection of atomic lines from VALD3 \citep{2015PhyS...90e4005R} together with roughly 15 million molecular lines representing 18 different molecules, the most important of which for this work are CaH (Plez, priv. comm.), MgH \citep{2003ApJS..148..599S,1995ASPC...78..205K}, and TiO \citep[with updates via VALD3]{1998A&A...337..495P}.

We use this grid to generate two synthetic libraries for radial velocity determination and parameter estimation.
For the WiFeS spectra, we use a coarsely sampled version of this grid, broadened to $R\sim7000$ with $5400 \leq \lambda \leq 7000$, $3000 \leq T_{\rm eff} \leq 8000\,$K, $3.0 \leq \log g \leq 5.5$, and $-1.0 \leq $[Fe/H]$ \leq 0.5$, in steps of $100\,$K, $0.25\,$dex, and $0.25\,$dex respectively. 

For the Echelle spectra, we adopted R=24,000 for $3000 \leq T_{\rm eff} \leq 6000\,$K, $4 \leq \log g \leq 5$, and  [Fe/H]=0, in steps of $250\,$K and $0.5\,$dex, respectively. Additionally, $\log g$ for $T_{\rm eff}<4000\,$K was extended to 5.5. Spectra cover wavelengths from 4800 to 6700~\AA.

\subsection{Radial velocities}
Radial velocities for datasets from both instruments were determined with the same algorithms using synthetic spectra described in the previous section.
\subsubsection{WiFeS}
Radial velocities of the WiFeS R7000 spectra were determined from a least squares minimisation of a set of synthetic template spectra varying in temperature (see Section \ref{sec.synthetic} for details of model grid). We use a coarsely sampled version of this grid, computed at R$\sim7000$ over $5400 \leq \lambda \leq 7000$ for $3000 \leq T_{\rm eff} \leq 5500\,$K, $\log g = 4.5$, and [Fe/H]$= 0.0$, with $T_{\rm eff}$ steps of $100\,$K for radial velocity determination. 

Prior to computing radial velocities, we normalise both our observed and synthetic template spectra. For warmer stars without the extensive molecular bands and opacities present in cool stars, continuum regions are typically used to continuum normalise the spectrum. For observed cool star spectra however, such regions are unavailable in the optical, so we must opt for another normalisation formalism, which we term here \textit{internally consistent normalisation}:
\begin{equation}
    f_{\rm norm} = f_{*} \times e^{\big(a_0 + \frac{a_1}{\lambda}+\frac{a_2}{\lambda^2}\big)} 
\end{equation}
where $f_{\rm norm}$ is the internally consistent normalised flux vector, $\lambda$ is the corresponding wavelength vector, and $a_0$, $a_1$, and $a_2$ are coefficients of a second order polynomial fitted to the logarithm of $f_{*}$, which is either an observed flux corrected spectrum, or a synthetic template. This functional form of normalisation has chosen to be largely independent of reddening.

Once generated, a given synthetic template (initially in the rest frame) can be interpolated and shifted to the science velocity frame as follows:
\begin{equation}
    f_{\rm temp,~rvs} = f_{\rm t}\big[\lambda \times \big(1-\frac{v_r-v_b}{c}\big)\big]
\end{equation}
where $f_{\rm temp,~rvs}$ is the RV shifted normalised template flux, $f_{\rm temp}$ is the template flux in the rest frame, $v_r$ and $v_b$ the radial and barycentric velocities respectively, and $c$ is the speed of light. $v_b$ is computed using the \texttt{ASTROPY} package \citep{astropy:2018} in \texttt{PYTHON}. 

Given a grid of $k$ different synthetic template spectra, the final radial velocity value is found by finding the synthetic template that best minimises:
\begin{equation}
    R(v_r) = \displaystyle\sum_{j}^{N}\bigg(\frac{{f_{\rm obs,~j} - f_{\rm temp,~rvs,~k,~j}(v_r)}}{\sigma_{{f_{\rm obs,~j}}}}\bigg)^2 M_j
\end{equation}
where $R$ is the total squared residuals as a function of radial velocity offset, $j$ is the pixel index, $N$ the total number of spectral pixels, $f_{\rm obs,~j}$ is the normalised observed flux at pixel $j$, $\sigma_{{f_{\rm obs,~j}}}$ is the uncertainty on $f_{\rm obs,~j}$, and $M_j$ is a masking term set to either 0 or 1 for each pixel. This step is done twice for each template spectrum, initially masking out only pixels affected by telluric contamination (H$_2$O: 6270-6290$\,$\SI{}{\angstrom}, and O2: 6856-6956$\,$\SI{}{\angstrom}), but then additionally masking out further pixels with high fit residuals. This second mask has the effect of excluding any pixels likely to skew the fit such as science target emission not present in the synthetic template (such as H$\alpha$). 

Least squares minimisation was done using the leastsq function from \texttt{PYTHON}'s \texttt{SCIPY} library, implemented in the \texttt{PYTHON} package \texttt{plumage}\footnote{\url{https://github.com/adrains/plumage}}. Statistical uncertainties on this approach are on average 430$\,$m$\,$s$^{-1}$, however per the work of \citet{2018MNRAS.480.5099K} we add this in quadrature with an additional 3$\,$km$\,$s$^{-1}$ uncertainty to account for WiFeS varying on shorter timescales than our hourly arcs can account for, and effects of variable star alignment on the slitlets. Note however that we do not employ corrections based on oxygen B-band absorption, demonstrated by \citealt{2018MNRAS.480.5099K} to improve precision, as such additional precision is unnecessary for this work and is difficult for cooler stars.

\begin{figure}
\includegraphics[width=\linewidth]{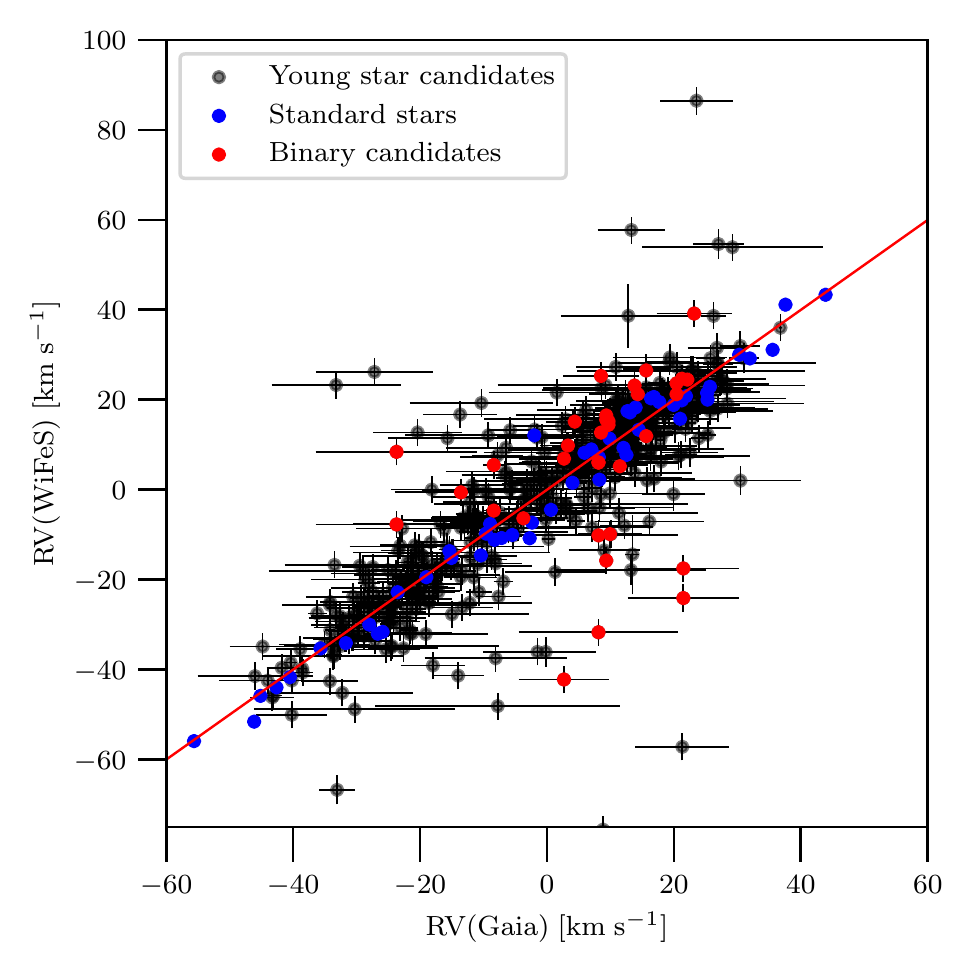}
\caption{A comparison between radial velocities from Gaia and from our pipeline for the WiFeS spectra. Standard stars (blue) have high S/N and small uncertainties. Binary star candidates (stars with repeated observations that show standard deviation of radial velocities greater than $5\,\mathrm{km\,s^{-1}}$ and stars that were classified as binaries by visual inspection) are marked in red.
}
\label{fig.rvs}
\end{figure}

Comparison of radial velocities for cool dwarf standard stars (e.g. from \citealp{2015ApJ...804...64M} and \citealp{2012ApJ...748...93R}, observed with the same instrument setup as part of Rains et al. in prep) with the Gaia catalogue \citep{2018A&A...616A...6S} shows an offset of WiFeS values for -1.7~$\mathrm{km\,s^{-1}}$ and a standard deviation of 3.2~$\mathrm{km\,s^{-1}}$ (Figure \ref{fig.rvs}). We suspect that most of the outliers are binary stars. Some of them are confirmed by either visual inspection or significally different radial velocities in case of repeated observations while there is not enough information available to investigate the rest of the interlopers.

\begin{figure}
\includegraphics[width=\linewidth]{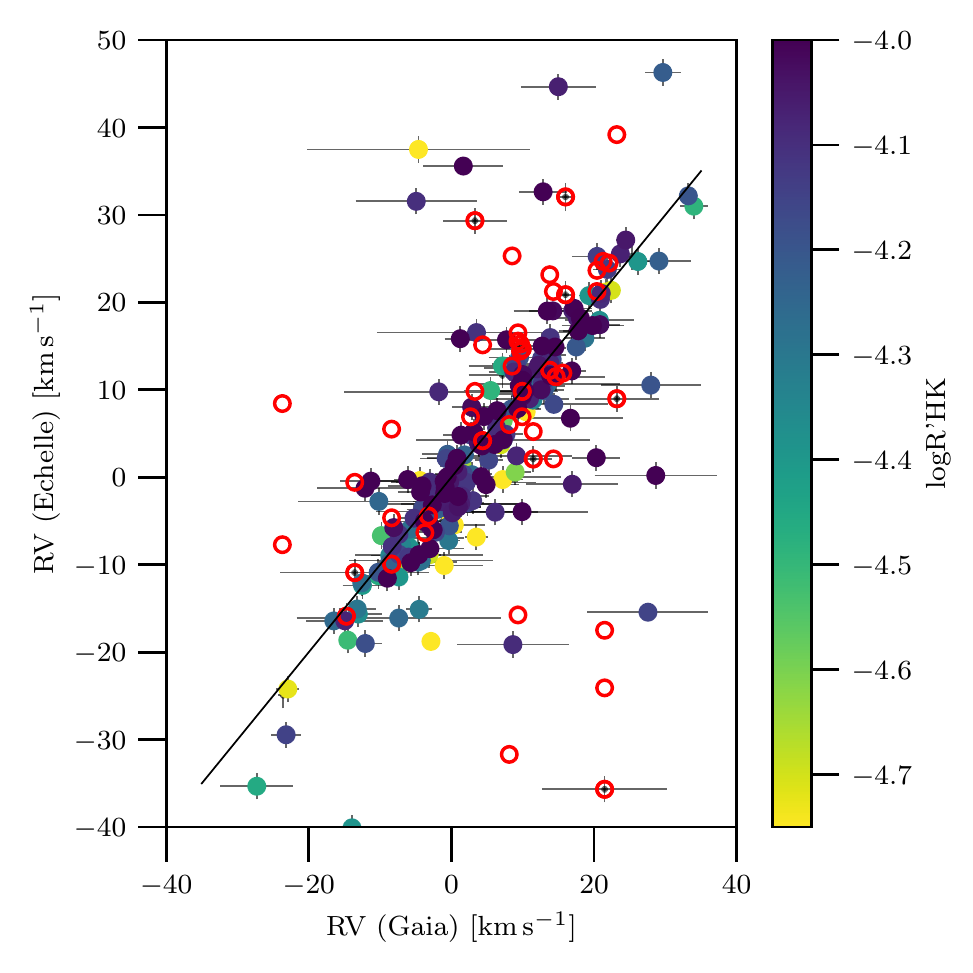}
\caption{A comparison between radial velocities from Gaia and our Echelle pipeline. Stars with the biggest disagreement with Gaia appear to be binary star candidates (red circles) or active (measured by calcium~II~H\&K emission $\mathrm{\log{R'_{HK}}}$, see Section \ref{sec.calcium}). The match with best-fitting template has been visually inspected for all stars in the sample.
}
\label{fig.rvs_echelle}
\end{figure}

\subsubsection{Echelle}
The same routine was utilized for the Echelle spectra on wavelengths between 5000 and 6500~\AA~ using their own synthetic library described in Sec. \ref{sec.synthetic}. As the correction for the blaze function and flux calibration were not performed in the data reduction step, each order within the relevant wavelength range was continuum normalized with a low order polynomial. Orders were then combined together into one spectrum in the range between 5000 and 6500~\AA. To match the continua of measured and synthetic libraries, fluxed model spectra were cut into wavelengths corresponding to Echelle orders, normalized and stitched back together with the same procedure. Finally, synthetic spectra were scaled to match 90th percentile of Echelle continua.

All spectra were visually inspected for any major reduction issues or other sources of peculiarity. Obvious double-lined binaries were flagged and their radial velocities are not reported in this work. Binary detection is reported in a separate column in Table \ref{tab.results}. 

Median internal uncertainty of derived radial velocity is 0.06~$\mathrm{km\,s^{-1}}$, but a combination of the systematic uncertainty and radial velocity jitter characteristic to young stars account for 1.5~$\mathrm{km\,s^{-1}}$.



Most of the stars have radial velocities consistent with Gaia (Figure \ref{fig.rvs_echelle}). Mean absolute deviation for stars with difference less than 10$\mathrm{km\,s^{-1}}$ is 0.6$\mathrm{km\,s^{-1}}$. There are a handful of outliers, and they all have large uncertainties in Gaia values. Some of those appear to be binary stars discovered either by visual inspection or large radial velocity difference in case of the repeated measurements. At the same time, a lost of such stars show high activity level (depicted by a measure of activity in calcium HK lines) that might dominate Gaia's calcium infrared triplet region used to determine radial velocities and cause systematic offsets. All Echelle stars have been visually inspected for possible peculiarity and their match with the best-fitting template.




\subsection{Reddening} \label{sec.reddening}
The M dwarf candidates are too close to be significantly reddened (<200~pc), but on the other hand they could remain embedded in their birth cocoons. At the same time, the sample is contaminated with hotter stars that lie in regions of more heavy extinction within the Galactic plane.
To derive an estimate for the intrinsic colour index (BP-W1)$_0$, temperatures of the best-matching templates were used as an input in the colour-temperature relation derived from the synthetic spectral library. Although Solar values were used to calibrate the zero point, a degree of uncertainties remains (increasing with colour) and the relation is only approximate. 
The resulting E(BP-W1) reveals a number of interlopers with temperatures higher than 4500~K. In particular, 156 WiFeS stars have E(BP-W1)$>$1 (20\% of the entire sample). 

\begin{figure}
\includegraphics[width=\columnwidth]{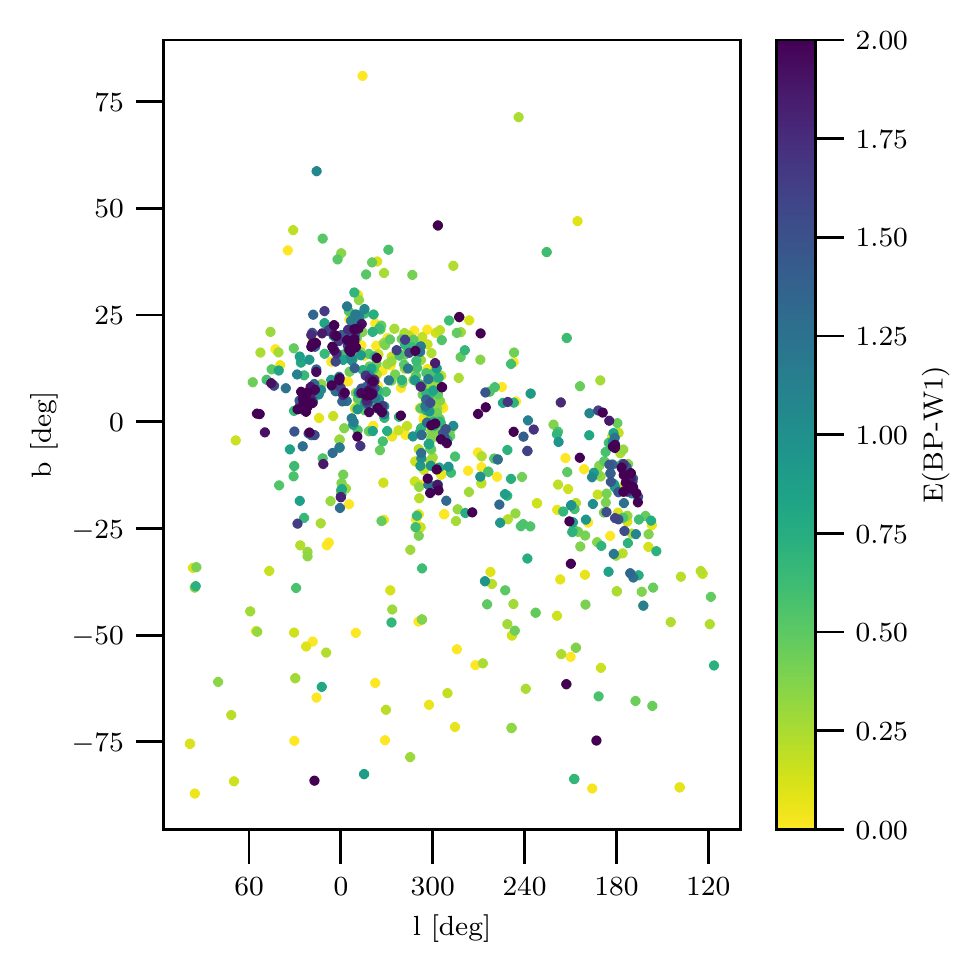}
\caption{The distribution of young candidates in the Galaxy. The majority of stars is found in clumps suggesting that they still reside close to their birth sites. The biggest group is found in the direction of the Scorpius-Centaurus OB2 region ($l>280$~deg). The second clump is likely the Hyades stars ($l\sim180$~deg). Colours show the interstellar reddening E(BP-W1).} 
\label{fig.galaxy}
\end{figure}

The estimated reddening E(BP-W1) is presented in Figure \ref{fig.cmd} together with the reddening vector\footnote{Reddening vector is determined for $A_V = 1$ and $R_V = 3.1$ from the \citet{1989ApJ...345..245C} model - \texttt{ccm89} in \url{https://extinction.readthedocs.io/en/latest/index.html}.}. Most interlopers with high reddening are found in the two regions in the Galactic plane with the highest concentration of stars in our sample: the Hyades and the Scorpius-Centaurus OB2 region (Fig. \ref{fig.galaxy}). Further analysis revealed that these stars do not show signs of youth and are likely located behind the local dust clouds associated with star-forming regions.

\section{Youth indicators} \label{sec.youth_indicators}
The following subsections address the characterization of the lithium absorption line and the excess emission in H$\alpha$ and Ca~II~H\&K lines for stars in our sample. A combination of all three values provides a robust indicator of the stellar youth. Algorithms used to measure the strengths of lithium and H$\alpha$ lines in this work are similar for data from both instruments WiFeS and Echelle. Excess emission in calcium is measured differently for Echelle due to low signal in the blue. 
All spectra, except the WiFeS calcium region, were locally normalized so that the youth features are surrounded by the continuum at 1 (and pseudo-continuum in M dwarfs). Binaries were not treated separately in this work and we provide youth indicators regardless of stars' multiplicity. All spectra were visually inspected for multiplicity and high rotation rate. We flag such cases in the final table and emphasize that this is qualitative inspection only and it is not complete. 

\subsection{Lithium}

The primary and most reliable spectroscopic feature sensitive to the age of the pre-main sequence dwarfs in the temperature range observed in our sample is the lithium 6708~\AA~ line. This absorption line is observed in low-mass pre-main sequence stars before the ignition of lithium in their interiors. Since these stars appear to be fully convective before their onset on the main sequence, the depletion of lithium throughout the entire star occurs almost instantly. Lithium is observed in F, G and early K dwarfs for up to $\sim$100~Myr (mass dependent), but late K and early M-type dwarfs deplete lithium much faster. For further information see \cite{2014prpl.conf..219S} and references therein. Both data and theoretical predictions show that at the age between $\sim$15-40~Myr there is practically no lithium left in these stars \citep{2015A&A...577A..42B, 2019MNRAS.484.4591Z}.


The strength of the lithium absorption lines in this work was characterized with the equivalent widths measured within 6707.8$\pm$1.4~\AA. 
Our spectra were pseudo-continuum normalized with a second order polynomial between 6700 and 6711~\AA. The lithium line was excluded from the continuum fit. The equivalent width was defined to be positive for lines in the absorption and was measured from the continuum level of 1.

In contrast to the emission-related features superimposed on the photospheric spectrum, the lithium absorption line shows a certain degree of correlation with the stellar rotation rate, e.g. \citet{2018A&A...613A..63B}. Fast rotators found by visual inspection are flagged in the table with results. While it appears to be fairly insensitive to the chromospheric activity \citep[e.g.][]{2019MNRAS.490L..86Y} 
it might in some cases be affected by strong veiling present in the classical T~Tauri stars \citep{1989AJ.....98.1444S}. Veiling is an extra source of continuum that causes absorption lines to appear weaker \citep{1990ApJ...363..654B}. However, measurements of H$\alpha$ emission described below reveal that no classical T~Tauri stars are present in the sample. Figure \ref{fig.youth_indicators} confirms a robust correlation between all three measures of the youth.

The distribution of EW(Li) shows a concentration of stars below 0.05~\AA, though we only consider positive detections in stars with values above this level. 
Repeated observations (45 stars) show 0.02~\AA~ of variation between individual measurements of the same object. 

\subsection{Calcium~II~H\&K} \label{sec.calcium}
It has long been known that atmospheric features associated with stellar activity in solar-like dwarfs anticorrelate with their age \citep{1972ApJ...171..565S, 1991ApJ...375..722S}. Empirical relations derived from chromospheric activity proxies enable age estimation of stars between $\sim$0.6-4.5~Gyr to a precision of $\sim$0.2~dex \citep{2008ApJ...687.1264M}. However, a combination of saturation \citep{2010ApJ...709..332B} and high variability \citep{1995ApJ...438..269B} of activity in younger stars prevents this technique yielding reliable results before the age of $\sim$200~Myr.
Nevertheless, a detection of a strong excess emission in the calcium~II~H\&K lines (Ca~II~H\&K; 3968.47 and 3933.66~\AA, respectively) -- a proxy for chromospheric activity -- helps to distinguish between active young stars and older stars with significantly lower emission rates.

A commonly used measure of stellar activity in solar-type stars is S-index introduced by \citealp{1978PASP...90..267V} and derived as

\begin{equation}
S = \alpha \frac{N_H + N_K}{N_V + N_R},
\end{equation}

where $N_H$ and $N_K$ are the count rates in a bandpass with a width of 1.09~\AA~ in the center of the Ca~II~H and K line, respectively. To match the definition of the first measurements obtained by a spectrometer at Mount Wilson Observatory \citep{1978ApJ...226..379W} and make the measurements directly comparable, counts are adjusted to the triangular instrumental profile as described in \citealp{1978PASP...90..267V}.
$N_V$ and $N_R$ are the count rates in 20~\AA-wide continuum bands outside the lines, centered at 3901.07~\AA~ and 4001.07~\AA. 

Constant $\alpha$ is a calibration factor that accounts for different instrument sensitivity and is derived by a comparison with literature S values. For WiFeS we provide a linear relation that converts measured S value on a scale directly comparable with the literature. For derivation see Appendix \ref{sec.appendix_s_index}.

Since $N_V$+$N_R$ has a color term due to nearby continuum shape varying with temperature, and because $N_H + N_K$ accounts for both chromospheric and photospheric contribution, it is more convenient to use the $\mathrm{R'_{HK}}$ index (first introduced by \citealp{1979ApJS...41...47L}) that represents a ratio between the chromospheric and bolometric flux and enables a direct comparison of activity in different stellar types. Using the conversion factor $C_{cf}$ that describes the colour-dependent relation between the S-index and the total flux emitted in the calcium lines, and $\mathrm{R_{phot}}$ that removes the photospheric contribution from the total flux in calcium, $\mathrm{R'_{HK}}$ is obtained as

\begin{equation}
    \mathrm{R'_{HK} = R_{HK} - R_{phot}}
\end{equation}

where $\mathrm{R_{HK}} = 1.887 \times 10^{-4}  \times  C_{cf}  \times  S$. The constant in the equation is taken from  \citealp{2017A&A...600A..13A}.
\citealp{1982A&A...107...31M} and \citealp{1984A&A...130..353R} provide the calibration of $C_{cf}$ and \citealp{1984ApJ...279..763N} and \citealp{1984ApJ...276..254H} for $\mathrm{R_{HK}}$ for the main sequence stars, but their relations become increasingly uncertain above B-V$>$1.2. \citealp{2017A&A...600A..13A} have recently extended the relation to M6 dwarfs (B-V$\sim$1.9) using HARPS data and calibrated the relation for colours that are more suitable for cool stars:
\begin{align}
    \log_{10}{C_{cf}} & = - 0.005c^3 + 0.071c^2 - 0.713c + 0.973 \\ 
    \log_{10}{R_{phot}} & = - 0.003c^3 + 0.069c^2 - 0.717c - 3.498. 
\end{align}

where $c=$V-K was determined from a low-order polynomial fit to the relation between synthetic BP-RP and V-K from \citealp{2018MNRAS.479L.102C}.

There are 26 stars in the sample with repeated observations. In general more active stars show higher variability rates. We provide a median value of 1.1$ \times 10^{-5}$ for the $\mathrm{R'_{HK}}$ variability. Stars with low levels of activity have measured $\log{\mathrm{R'_{HK}}}=$ -5 or lower and we consider them inactive.

Activity in the Echelle spectra was evaluated in the same way as WiFeS stars. The calibration of the S-index was done using 19 stars observed with both instruments. For more details on the calibration see Appendix \ref{sec.appendix_s_index}.

\begin{figure}
\includegraphics[width=\columnwidth]{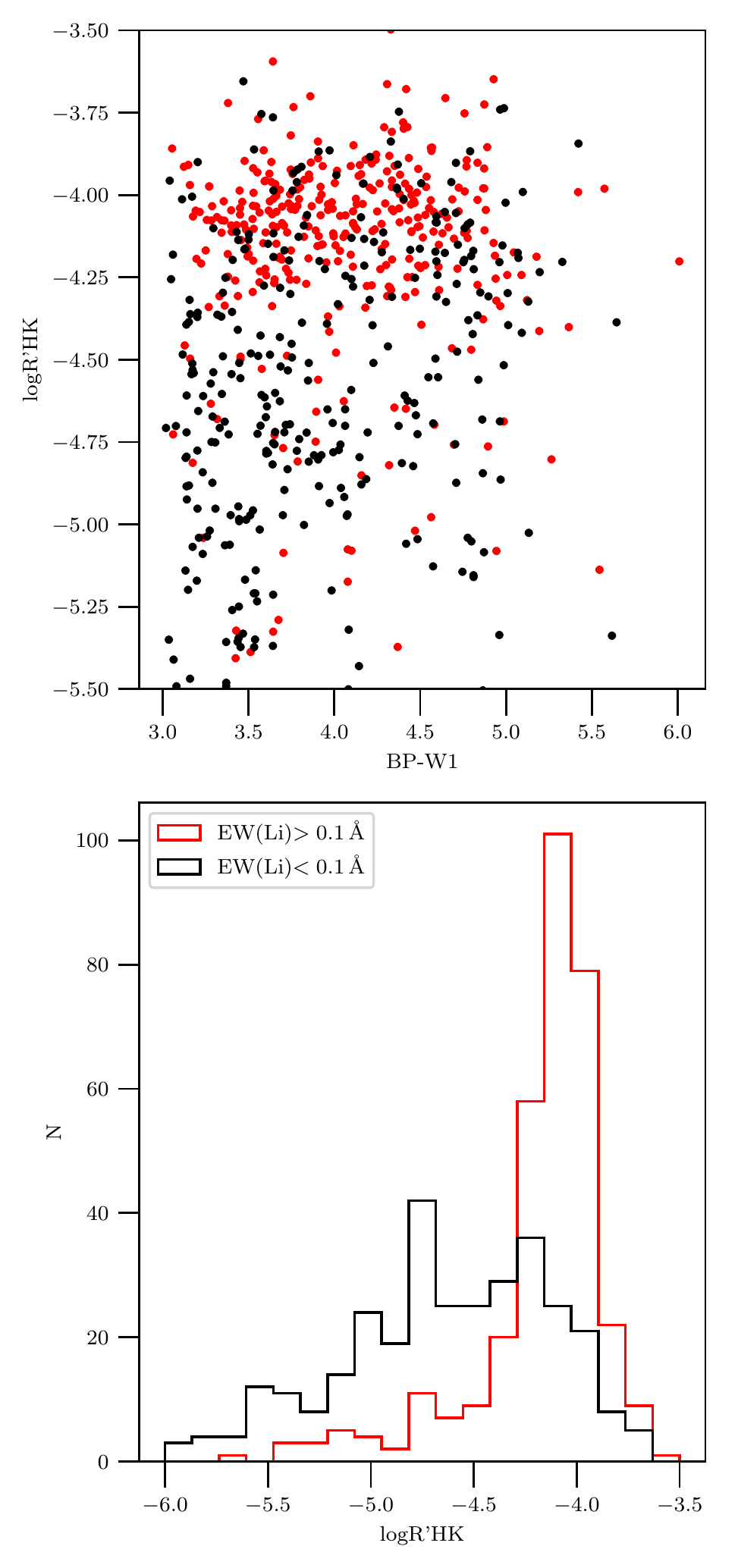}
\caption{\textit{Upper panel:} The introduction of the $\mathrm{logR'_{HK}}$ index minimises the color term and allows for comparison of activity rates among different spectral types. \textit{Lower panel:} Distribution of $\mathrm{logR'_{HK}}$ index for 680 stars. Nearly all stars with a detectable lithium show very strong calcium emission.
}
\label{fig.rhk}
\end{figure}

The distribution of $\log{\mathrm{R'_{HK}}}$ is known to be bimodal for the main sequence stars in the Solar neighbourhood (e.g. \citealp{2003AJ....126.2048G}). Figure \ref{fig.rhk} shows two peaks, but they are centered at higher levels of activity due to our focus on the pre-main sequence stars. The more active peak is found at $\sim -4$ where $\log{\mathrm{R'_{HK}}}$ saturates for stars with rotation rates less than 10~days \citep{2017A&A...600A..13A}. According to \citealp{2008ApJ...687.1264M}, such high activity levels occur at ages of $\sim$10~Myr. We also plot $\log{\mathrm{R'_{HK}}}$ versus colour (the same figure) to confirm that the colour term is minimized.

\begin{figure*}
\includegraphics[width=\linewidth]{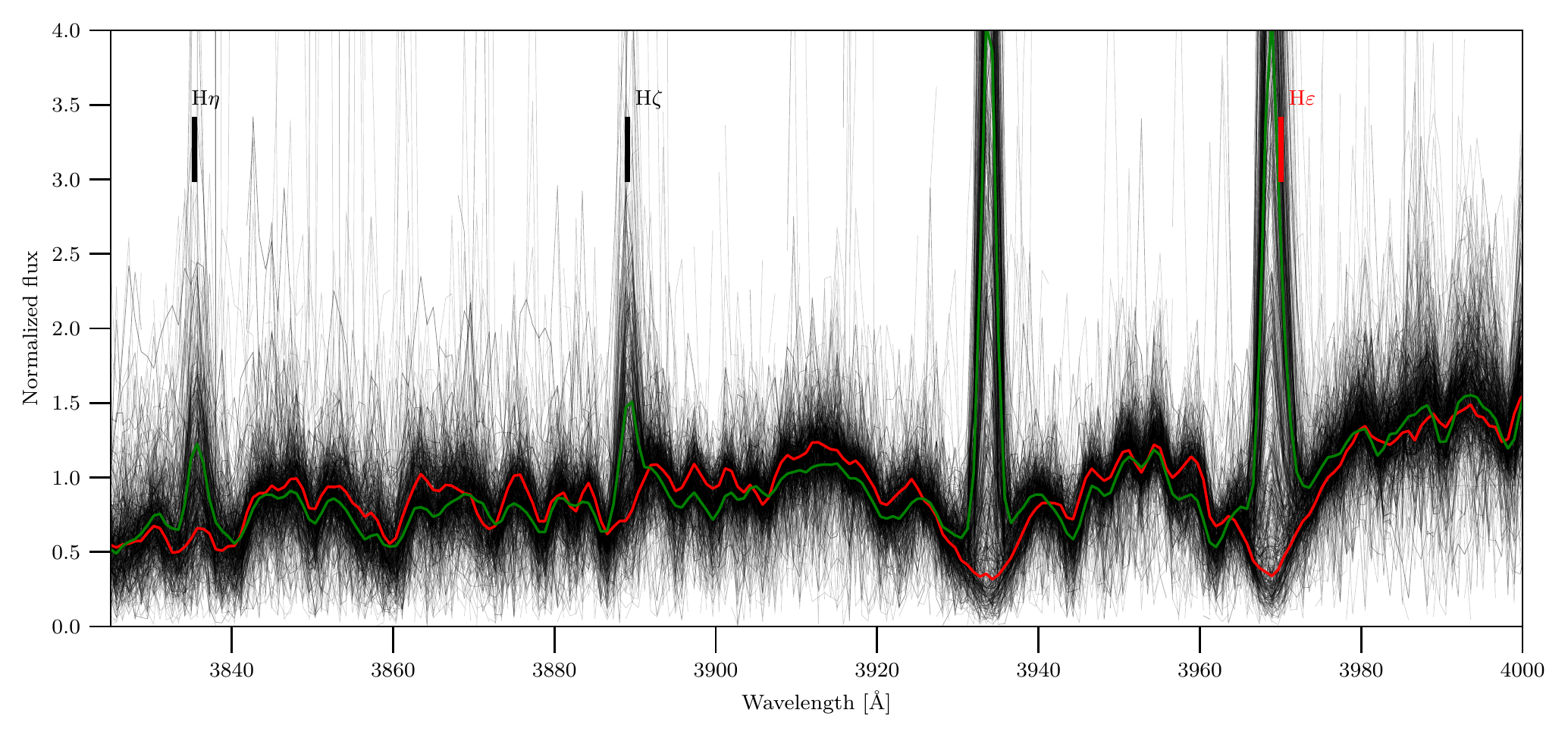}
\caption{Calcium lines in the WiFeS spectra. Ca~II~H line appears to be wider than Ca~II~K due to the presence of the Balmer emission line at 3970~\AA. Red spectrum is a median spectrum with $\mathrm{logR'_{HK}}<$-4.9. Very active spectra with $\mathrm{logR'_{HK}}>$-4.4 (green) are young and show Balmer emission.
}
\label{fig.wifes_calcium_balmer}
\end{figure*}

There are two sets of lines that cause strong emission in this wavelength range: calcium~II~H\&K lines and Balmer emission lines in the youngest stars. Calcium H line is in some cases strongly blended by the Balmer emission line in the WiFeS spectra but count rate was measured within 1.09~\AA~ (see Fig. \ref{fig.wifes_calcium_balmer}). 

\subsection{Balmer series}
While weak and moderate excess emission rates in the H$\alpha$ line (6562.8~\AA) are associated with chromospheric activity  \cite[e.g.][]{2004AJ....128..426W, 2008AJ....135..785W}, strong emission in the entire Balmer series, with H$\alpha$ being especially prominent ($>$10~\AA), is typically observed in classical T~Tauri stars that are low-mass objects younger than $\sim$10~Myr \citep{1989ARA&A..27..351B, 1989A&ARv...1..291A, 1998AJ....115..351M, 2006MNRAS.370..580K, 2014prpl.conf..219S}. It is widely accepted that there is a tight correlation between the average chromospheric fluxes emitted by the Ca~II~H\&K and H$\alpha$ lines \cite[e.g.][]{1995A&A...294..165M}. Although \citet{2007A&A...469..309C} report that this relation is more complicated, emission in H$\alpha$ represents a robust indicator of stellar youth. 
Characterisation of stellar activity from the H$\alpha$ line is especially convenient in late-type dwarfs that only present a weak photosphere in the blue where Ca~II~H\&K are located. 



The equivalent width of H$\alpha$ lines was measured between 6555 and 6567~\AA~ relative to the continuum, e.g. (1~-~flux) in the H$\alpha$ region. Negative values thus indicate absorption while positive values denote emission above the continuum. Interpretation of these results is not straightforward due to a wide range of the H$\alpha$ line profiles being strongly affected not only by the temperature but also the surface gravity. However, most of the stars show strong emission that is in any case an indicator for extreme stellar youth. We make a conservative estimate and only treat spectra with EW(Ha)$>$-0.5~\AA~ as active (see Fig. \ref{fig.wifes_calcium_balmer}).
Repeated observations of 45 stars reveal a typical difference between the maximal and minimal EW(Ha) value of 0.2~\AA. This uncertainty might also include a variability component of stellar activity.


Based on equivalent widths of H$\alpha$, most of the stars with excess emission belong to either weak (EW(Ha)$<$5~\AA) or post-T~Tauri stars. One third of the entire sample shows emission in the entire Balmer series. Column \texttt{Balmer} in Table \ref{tab.results} lists objects with clear Balmer emission that was detected by visual inspection.



\section{Discussion} \label{sec.discussion}
A combination of the three complementary youth features -- excess emission in Ca~II~H\&K and H$\alpha$ associated with magnetic fields active but declining for billions of years, and lithium absorption line present for a few 10~Myr in late K and early M dwarfs -- maximises the estimated age range and the robustness of our young star identification.

\begin{figure}
\includegraphics[width=\columnwidth]{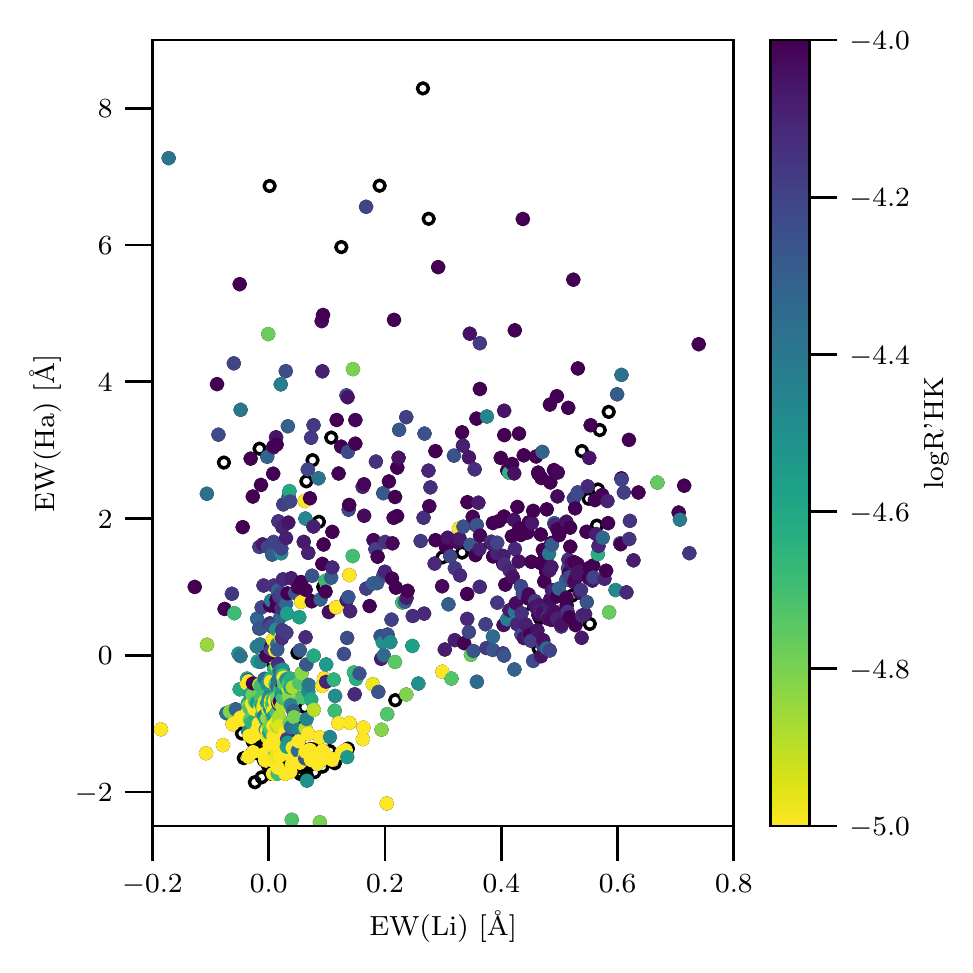}
\caption{Youth indicators studied in this work show a high degree of correlation. Chromospheric activity in young stars shows a high level of variability over time, but there appears to be a lower limit for H$\alpha$ emission with respect to the strength of the lithium line. Stars with no $\log{\mathrm{R'_{HK}}}$ available are marked with circles.
}
\label{fig.youth_indicators}
\end{figure}

This work uncovered 549 sources with at least one of the three indicators above the detection limit:
EW(Li)$>$0.1~\AA~ or EW(Ha)$>$-0.5~\AA~ or $\log{\mathrm{R'_{HK}}}>-4.75$. The strategy is thus 70\% successful. In particular, there are 281 stars with all three indicators above the detection limit.
There are 346 stars with a detectable lithium line (44\%), 479 with $\mathrm{EW(H}\alpha)>-0.5$ (60\% of the sample) and 464 objects (60\%) with a detectable calcium emission. Not surprisingly, there are 409 stars that show both calcium and H$\alpha$ youth features, as these two indicators are well correlated due to their common origin in chromospheric activity. 
The lithium absorption line undergoes a different mechanism (lithium depletion in the pre-main sequence phase) and is much more short-lived. This causes an overdensity of chromospherically active stars with high H$\alpha$ but no lithium left (Fig. \ref{fig.youth_indicators}).
There are 10 stars in the sample that display lithium absorption but show no chromospheric activity. 

The figure also shows that all stars with strong lithium  emit excess flux in their chromospheres. This explains the void in the bottom right part of this figure. 
Note that a small subset of individual stars only has one or two youth indicators measured due to noise in the respective spectral regions.

All youth indicators, radial velocities and flags denoting Balmer emission, binarity and fast rotation are listed in Table \ref{tab.results}, together with their 2MASS identifiers \citep{2003yCat.2246....0C}. 
Even though our selection avoided known young stars, we cross-matched our catalogue with the literature. We found 15 stars in common with the list of association members described by  \citet{2018ApJ...856...23G} and 6 from \citet{2018ApJ...860...43G}. We found 9 objects from our list in the work by \citet{2009A&A...508..833D} measuring lithium lines of $\sim$400 objects, and 3 overlapping stars with \citealp{2015MNRAS.448.2737R} who targeted stars from Upper Scorpius that were mostly fainter than our magnitude limit.
In total, 33 unique objects out of 766 from our list (4\%) are known association members or have lithium measured in the literature, and the rest are considered new detections.



\begin{figure}
\includegraphics[width=\columnwidth]{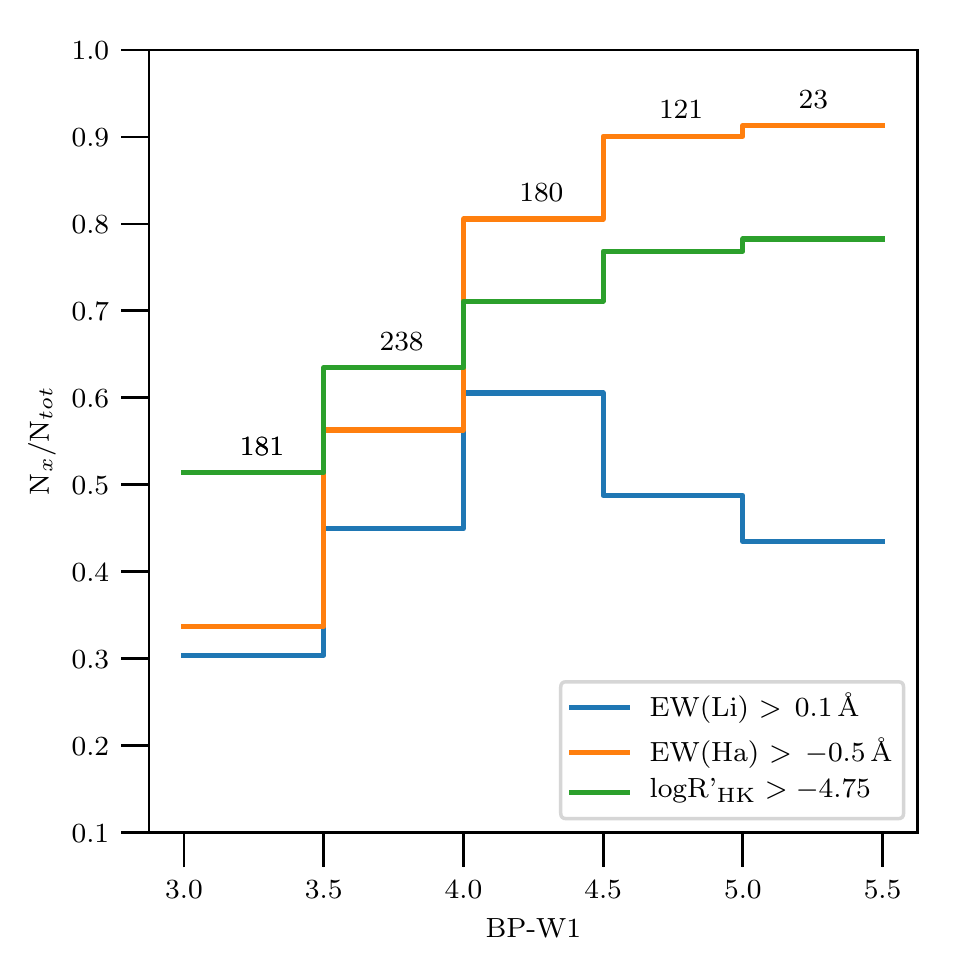}
\caption{Strategy success as a fraction of young stars with detectable spectroscopic features of youth versus their color. Detection rate for calcium and H$\alpha$ increase towards redder stars with different slopes. This might be due to a dependence of EW(H$\alpha$) on the temperature. Lithium absorption line is observed only in the youngest stars. Detection rate drops for early M dwarfs because they deplete lithium the fastest.
The number of all candidates in each colour bin is shown in the plot.
}
\label{fig.strategy_success}
\end{figure}

The occurrence rate for all youth features is color dependent (Fig. \ref{fig.strategy_success}). Cooler stars in general more likely show signs of youth. Due to their slower evolution they spend more time above the main sequence and display signs of their youth much longer. However, we observe a drop in the occurrence rate of the lithium line in M dwarfs. This is because they deplete lithium the fastest and soon fall below the detection limit. 

\begin{figure*}
\includegraphics[width=\linewidth]{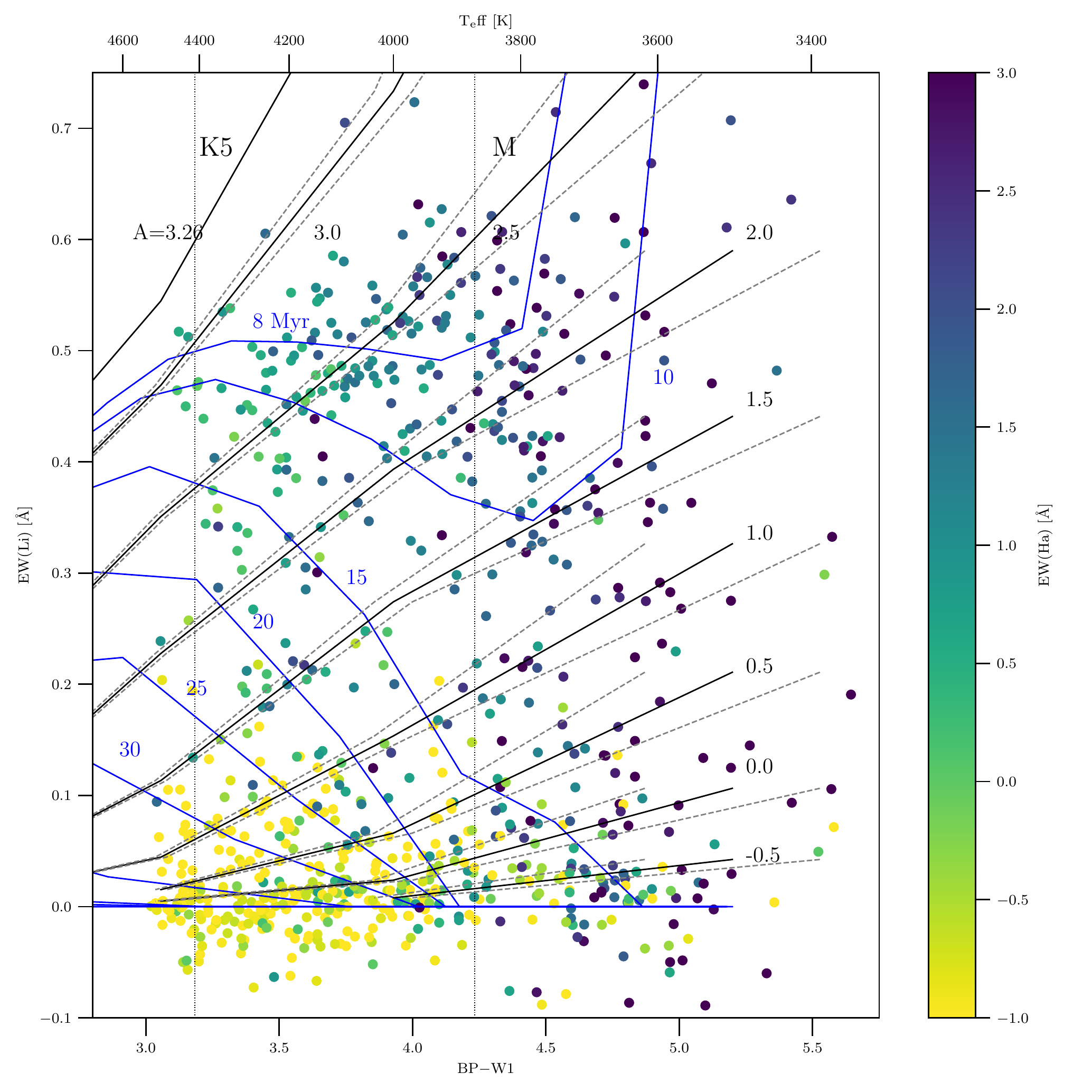}
\caption{Lithium isochrones (blue lines) reveal a number of very young stars in the sample ($<$25~Myr). 349 stars have a detectable lithium with EW(Li)$>$0.1~\AA. Black lines show lithium abundances with their uncertainties (dashed). Lithium strength correlates well with the excess emission in the H$\alpha$ line.
}

\label{fig.isochrones}
\end{figure*}

Lithium isochrones enable age estimation for late K and early M dwarfs younger than 15-40~Myr. We follow \citet{2019MNRAS.484.4591Z} and take indicative non-LTE equivalent widths from \citet{1996A&A...311..961P} for Solar metallicity and $\log{g} = 4.5$. We combine them with the \citet{2015A&A...577A..42B} models of lithium depletion (assuming the initial absolute abundance of 3.26 from \citealp{2009ARA&A..47..481A}) to compute lithium isochrones (Fig. \ref{fig.isochrones}).
Lines indicating abundances in the plot show that EW(Li) in our temperature range practically traces any amount of lithium left in the atmosphere.

There appears to be an overdensity of 278 objects above EW(Li)$>$0.3~\AA~ corresponding to the ages of 15~Myr and younger. Moreover, there are 325 stars lying above the 20~Myr isochrone and the 0.1~\AA~ detection limit.
Figure \ref{fig.youth_indicators} confirms that stars with the strongest lithium have the highest $\log{\mathrm{R'_{HK}}}$ values of -4 which corresponds to $\sim$10~Myr according to the \citealp{2008ApJ...687.1264M} activity-age relation.
These objects likely belong to the Scorpius-Centaurus association -- especially because their $(l,b)$ location overlaps with this region in the sky. However, further kinematic analysis is needed to confirm their membership.

Since our selection encompass all stars above the main sequence, the sample is contaminated with stars with bad astrometric solutions. 45\% of our observed objects have \textit{re-normalised unit weight error} (the \texttt{RUWE} parameter from the Gaia~DR2 tables describing the goodness of fit to the astrometric observations for a single star) greater than 1.4. Gaia~DR2 documentation suggests that such stars either have a companion or their astrometric solution is problematic. There is no detectable lithium left in these stars and they appear to be old in our context with low or zero emission in calcium and H$\alpha$. When stars with \texttt{ruwe}$>$1.4 and high reddening are removed from our catalog, 80\% of stars left show at least one spectroscopic sign of stellar youth. This suggests a high efficiency in selection of young stars from the Gaia catalog based on their overluminosity and a reliable astrometric single star solution.

\section{Conclusions} \label{sec.conclusions}
We selected and observed 766 overluminous late K and early M dwarfs with at least 1~magnitude above the main sequence and with Gaia G magnitude between 12.5 and 14.5. The kinematic cut was wide enough to avoid a bias towards higher-mass stars and include low-mass dwarfs. Observations were carried out over 64 nights with the Echelle and Wide Field Spectrographs at the ANU 2.3m telescope in Siding Spring observatory. 
The analysis revealed 544 stars with at least one feature of stellar youth, i.e. the lithium absorption line or excess emission in H$\alpha$ or calcium~H\&K lines. The strength of the lithium absorption line indicates that 349 stars are younger than 25~Myr.

This sample significantly expands the census of nearby young stars and adds 512 new young stars to the list. Only 33 out of 544 objects with at least one youth indicator are listed in external catalogs of young stars. For example, \citealp{2018ApJ...856...23G} characterised known nearby associations and provided a list of 1400 young stars from a wide variety of sources. Our catalog has only 15 stars in common with theirs and thus expands the sample by 35\%. Although a further kinematic analysis is needed to confirm their membership, it is likely that a great fraction of stars from our sample belong to the Scorpius-Centaurus association because they are found in that direction in the sky and all have lithium ages $<$20~Myr. However, we only find 3 stars in common with \citealp{2015MNRAS.448.2737R} who kinematically and photometrically selected and observed mostly fainter stars in Upper-Scorpius.
Strong lithium absorption lines and excess emission in calcium in these objects consistently indicate likely stellar ages of roughly 10~Myr, according to the activity--age relation \citep{2008ApJ...687.1264M} and lithium isochrones (see Fig. \ref{fig.isochrones}). The latter reveal 325 stars with $\mathrm{EW(Li)>0.1}$~\AA~ and above the 20~Myr isochrone. 


We report on a high success rate in search for young stars by selecting overluminous objects in the Gaia catalog. After stars with unreliable astrometry ($\mathrm{ruwe}>1.4$ that indicates bad astrometry or multiplicity) and high reddening are removed, the success rate is 80\%.




Radial velocities are determined for spectra from both instruments, with average uncertainties of 3.2~$\mathrm{km\,s^{-1}}$ for WiFeS and 1.5~$\mathrm{km\,s^{-1}}$ for Echelle stars. 
This catalog of nearby young stars now has all kinematic measurements available to improve the analysis of young associations and help to find their birthplace. For example, \citealp{2020MNRAS.499.5623Q} have recently shown that stellar associations come from different places in the Galaxy. Follow up work may include e.g. using Chronostar \citep{2019MNRAS.489.3625C} to provide kinematic ages, robust membership estimates and orbital models of young associations to infer the origins of this sample, as well as the extraction and analysis of rotational periods using TESS to obtain ages using gyrochronology where possible.

\section*{Acknowledgements}
We acknowledge the traditional owners of the land on which the telescope stands, the Gamilaraay people, and pay our respects to elders past and present. This work has made use of data from the European Space Agency (ESA) mission Gaia (https://www.cosmos.esa.int/gaia), processed by the Gaia Data Processing and Analysis Consortium (DPAC, https://www.cosmos.esa.int/web/gaia/dpac/consortium). Funding for the DPAC has been provided by national institutions, in particular the institutions participating in the Gaia Multilateral Agreement. This publication makes use of data products from the Wide-field Infrared Survey Explorer, which is a joint project of the University of California, Los Angeles, and the Jet Propulsion Laboratory/California Institute of Technology, funded by the National Aeronautics and Space Administration. M{\v Z} acknowledges funding from the Australian Research Council (grant DP170102233). ADR acknowledges support from the Australian Government Research Training Program, and the Research School of Astronomy \& Astrophysics top up scholarship. This research made use of Astropy, a community-developed core Python package for Astronomy (Astropy Collaboration 2013, 2018). 
Parts of this research were supported by the Australian Research Council Centre of Excellence for All Sky Astrophysics in 3 Dimensions (ASTRO 3D), through project number CE170100013.
Parts of this research were conducted by the Australian Research Council Centre of Excellence for Gravitational Wave Discovery (OzGrav), through project number CE170100004. S.-W. Chang acknowledges support from the National Research Foundation of Korea (NRF) grant, No. 2020R1A2C3011091, funded by the Korea government (MSIT).

Software: \texttt{numpy} \citep{harris2020array}, \texttt{scipy} \citep{2020SciPy-NMeth}, \texttt{ipython} \citep{doi:10.1109/MCSE.2007.53}, \texttt{pandas} \citep{mckinney-proc-scipy-2010}, \texttt{matplotlib}
\citep{doi:10.1109/MCSE.2007.55} and \texttt{astropy} \citep{astropy:2018}.

\section*{Data Availability}

This work is based on publicly available databases. Gaia data is available on \href{https://gea.esac.esa.int/archive/}{https://gea.esac.esa.int/archive/} together with the crossmatch with 2MASS and WISE catalogs.
A compilation of known young stars with S-indices from \citealp{2013A&A...551L...8P} is available on \href{http://vizier.u-strasbg.fr/viz-bin/VizieR?-source=J/A+A/551/L8&-to=3}{http://vizier.u-strasbg.fr/viz-bin/VizieR?-source=J/A+A/551/L8\&-to=3}.
All measurements from this work are provided in the appendix with a full table available online.




\bibliographystyle{mnras}
\bibliography{paper} 


\appendix

\section{Table with results}
\newcommand{\mytable}{\input{table.tex}}
\begin{landscape}
\begin{table*}
\hspace*{-8cm}
\begin{tabular}{llrrrlrlccrrrrrrrr}
Designation & 2MASS & G & BP-W1 & BP-RP & ruwe & obsdate & Inst. & S/N(B) & S/N(R) & RV & $\sigma_{\mathrm{RV}}$ & logR'HK & EW(Ha) & EW(Li) & Binary & Balmer \\ 
\textit{Gaia} DR2 & & & & & & & & & & $\mathrm{km\,s^{-1}}$ & $\mathrm{km\,s^{-1}}$ &  & $\mathrm{\mathring{A}}$ & $\mathrm{\mathring{A}}$ & & \\
 \hline
 \input{table.tex}
\end{tabular}
\caption{List of young star candidates with their radial velocities and youth signatures. Full table is available in the online version.}
\label{tab.results}
\end{table*}
\end{landscape}

\section{Calibration of S-index} \label{sec.appendix_s_index}
\subsection{WiFeS}
In order to calibrate the S-index measured with the WiFeS instrument ($\mathrm{S_{raw}}$) and bring it to the scale comparable with Mount Wilson index, a set of supplementary stars from the literature was observed (Rains et al., in prep.).
Table \ref{tab.calibrator_stars} lists 30 stars from \citealp{2013A&A...551L...8P} who combined data from many different sources. References are listed in the table, and we keep the notation from the original paper to avoid confusion and retain any extra information.

\begin{figure}
\includegraphics[width=\linewidth]{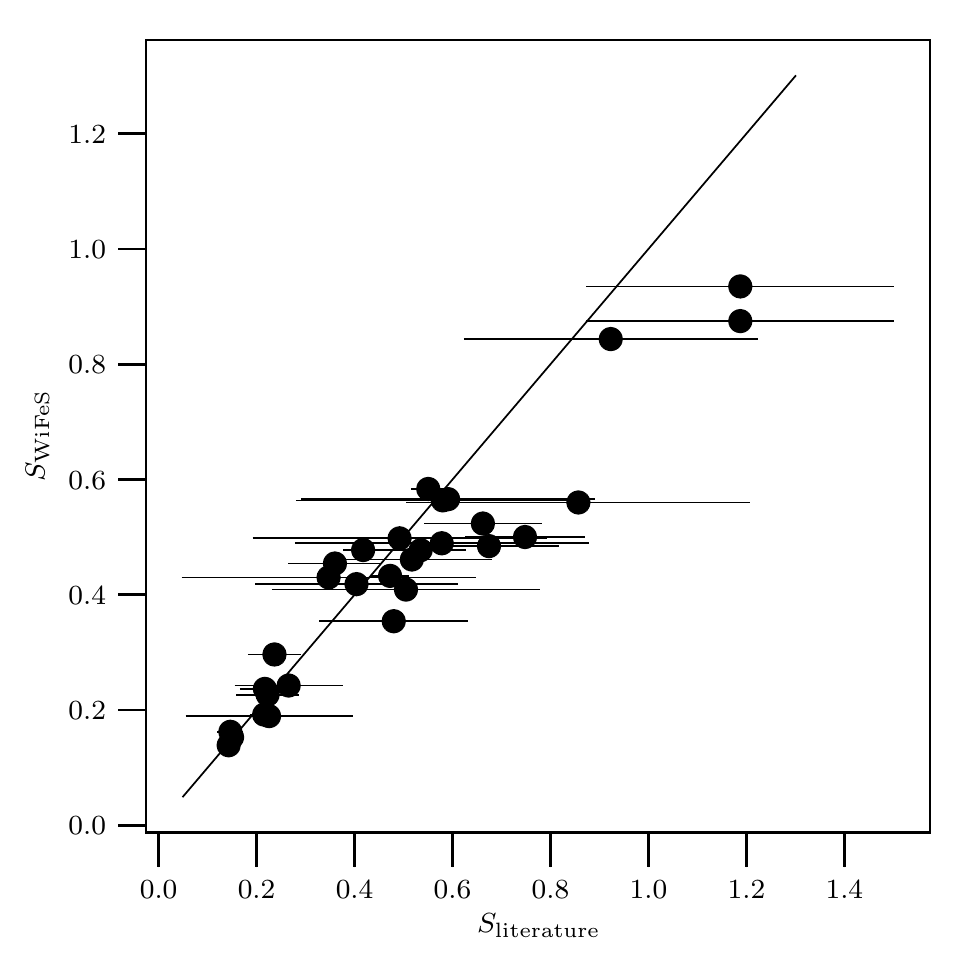}
\caption{Calibration of $\mathrm{S_{WiFeS}}$ with 30 stars from the literature. Errorbars are displayed for stars with repeated measurements and show the span of both measurements. The central value is an average and it is used in the fit.
}
\label{fig.calibrate_S_index}
\end{figure}

These selected stars cover the entire range of activity levels.
Due to high variability with time and stellar cycles, this catalog often reports $\mathrm{S_{min}}$ and $\mathrm{S_{max}}$. In such cases we take the median value and assign standard deviation as its uncertainty. A linear fit
\begin{equation}
   \mathrm{S_{WiFeS}} = 20.490 \times \mathrm{S_{raw\,WiFeS}} -0.112
\end{equation}
enables a fair reconstruction of the literature values (Figure \ref{fig.calibrate_S_index}). Note that uncertainty of this fit is rather large ($\sim$1 in the slope) due to variability of activity in some of the targets.

\begin{table*}
\begin{tabular}{rrrrrrrrrr}
HD & Gaia DR2 & obsdate & Sraw & Smin & Smax & logRmin & logRmax & BP-RP & Refs\\
 \hline
10700 & 2452378776434276992 & 20190722 & 0.015 & 0.055 & 0.396 & -6.311 & -4.385 & 1.00 & 12334557899aabbfhjj \\
32147 & 3211461469444773376 & 20200201 & 0.017 & 0.155 & 0.376 & -5.492 & -4.915 & 1.25 & 24556aaf \\
154363 & 4364527594192166400 & 20200201 & 0.026 & 0.197 & 0.611 & -5.473 & -4.817 & 1.47 & aaefjj \\
2151 & 4683897617108299136 & 20190722 & 0.013 & 0.120 & 0.173 & -5.283 & -4.864 & 0.78 & 33799 \\
36003 & 3210731015767419520 & 20191014 & 0.028 & 0.265 & 0.455 & -5.221 & -4.916 & 1.38 & 67aajj \\
10697 & 95652018353917056 & 20190826 & 0.012 & 0.128 & 0.158 & -5.178 & -4.982 & 0.87 & 55aajj \\
190248 & 6427464325637241728 & 20190826 & 0.013 & 0.131 & 0.169 & -5.173 & -4.952 & 1.07 & 33799f \\
103932 & 3487062064765702272 & 20200201 & 0.023 & 0.328 & 0.632 & -5.147 & -4.800 & 1.36 & 7aafjj \\
108564 & 3520548825260557312 & 20200203 & 0.015 & 0.186 & 0.245 & -5.142 & -4.962 & 1.25 & 7g \\
4628 & 2552925644460225152 & 20190826 & 0.017 & 0.159 & 0.286 & -5.124 & -4.737 & 1.11 & 23556799aaef \\
155203 & 5965222838404324736 & 20191017 & 0.020 & 0.182 & 0.291 & -5.024 & -4.500 & 0.70 & 37 \\
26965 & 3195919528988725120 & 20190722 & 0.017 & 0.166 & 0.268 & -5.016 & -4.698 & 1.03 & 234559aa \\
200779 & 1736838805468812160 & 20191015 & 0.029 & 0.531 & 0.818 & -4.991 & -4.776 & 1.51 & 56aajj \\
21197 & 5170039502144332800 & 20200203 & 0.030 & 0.626 & 0.870 & -4.924 & -4.763 & 1.39 & 6aajj \\
209100 & 6412595290592307840 & 20190826 & 0.028 & 0.354 & 0.680 & -4.914 & -4.578 & 1.28 & 4799 \\
156026 & 4109034455276324608 & 20200201 & 0.033 & 0.506 & 1.208 & -4.895 & -4.473 & 1.40 & 23455799aafjj \\
22049 & 5164707970261630080 & 20190722 & 0.025 & 0.231 & 0.779 & -4.838 & -4.192 & 1.12 & 23345567899aaf \\
101581 & 5378886891122066560 & 20200201 & 0.027 & 0.433 & 0.512 & -4.822 & -4.736 & 1.32 & 47f \\
50281 & 3101923001490347392 & 20200201 & 0.031 & 0.542 & 0.782 & -4.707 & -4.527 & 1.29 & 6aafjj \\
61606 & 3057712223051571200 & 20200201 & 0.029 & 0.443 & 0.627 & -4.647 & -4.472 & 1.15 & 6aafgjj \\
171825 & 6439391797712630784 & 20200912 & 0.030 & 0.492 & 0.492 & -4.593 & -4.593 & 1.18 & 7 \\
208272 & 6617495364101129728 & 20200912 & 0.026 & 0.347 & 0.347 & -4.588 & -4.588 & 1.02 & 7 \\
18168 & 5049234888291201280 & 20200912 & 0.034 & 0.516 & 0.585 & -4.569 & -4.506 & 1.17 & 17 \\
224789 & 4703237305086965376 & 20200912 & 0.029 & 0.377 & 0.458 & -4.557 & -4.455 & 1.04 & 17 \\
158866 & 5774205537990380160 & 20200912 & 0.033 & 0.591 & 0.591 & -4.548 & -4.548 & 1.18 & 7 \\
216803 & 6604147121141267712 & 20190722 & 0.051 & 0.873 & 1.502 & -4.541 & -4.288 & 1.33 & 4799aaijj \\
216803 & 6604147121141267712 & 20200912 & 0.048 & 0.873 & 1.502 & -4.541 & -4.288 & 1.33 & 4799aaijj \\
924 & 4996401292991097600 & 20200912 & 0.033 & 0.580 & 0.580 & -4.419 & -4.419 & 1.10 & 7 \\
9054 & 4916062039935185792 & 20200912 & 0.070 & 0.911 & 0.911 & -4.324 & -4.324 & 1.20 & 7 \\
6838 & 5034700237924390016 & 20200912 & 0.029 & 0.578 & 0.578 & -4.311 & -4.311 & 1.01 & 7 \\
223681 & 6530566531700652544 & 20200912 & 0.047 & 0.923 & 0.923 & -4.153 & -4.153 & 1.14 & 7 \\
\end{tabular}
\caption{List of stars from the \citealp{2013A&A...551L...8P} compilation with literature values and used here to calibrate the S-index. $\mathrm{S_{raw}}$ (observed on date '$\mathrm{obsdate}$') is measured in this work. 
References are listed in the original form from \citealp{2013A&A...551L...8P} (doubled letters correspond to studies with repeated measurements): (2) \citealp{1995ApJ...438..269B}, (3) \citealp{2008A&A...483..903B}, (4) \citealp{2007A&A...469..309C}, (5) \citealp{1991ApJS...76..383D}, (6) \citealp{2003AJ....126.2048G}, (7) \citealp{2006AJ....132..161G}, (8) \citealp{2007AJ....133..862H}, (9) \citealp{1996AJ....111..439H}, (a) \citealp{2010ApJ...725..875I}, 
(b) \citealp{2011A&A...531A...8J},
(e) \citealp{2010A&A...514A..97L},
(f) \citealp{2009A&A...493.1099S},
(g) \citealp{2000A&AS..142..275S},
(h) \citealp{2002MNRAS.332..759T},
(i) \citealp{2007AJ....133.2524W},
(j) \citealp{2004ApJS..152..261W}.
}
\label{tab.calibrator_stars}
\end{table*}


\subsection{Echelle}
Calibration of the Echelle S-index is based on stars that were observed with both instruments. We compare $\mathrm{S_{WiFeS}}$ with $\mathrm{S_{raw\,Echelle}}$ and determine a relation that converts 
$\mathrm{S_{raw\,Echelle}}$ to $\mathrm{S_{Echelle}}$:
\begin{equation}
    \mathrm{S_{Echelle}} = 0.473 \times \mathrm{S_{raw\,Echelle}} +0.830.
\end{equation}

\begin{figure}
\includegraphics[width=\linewidth]{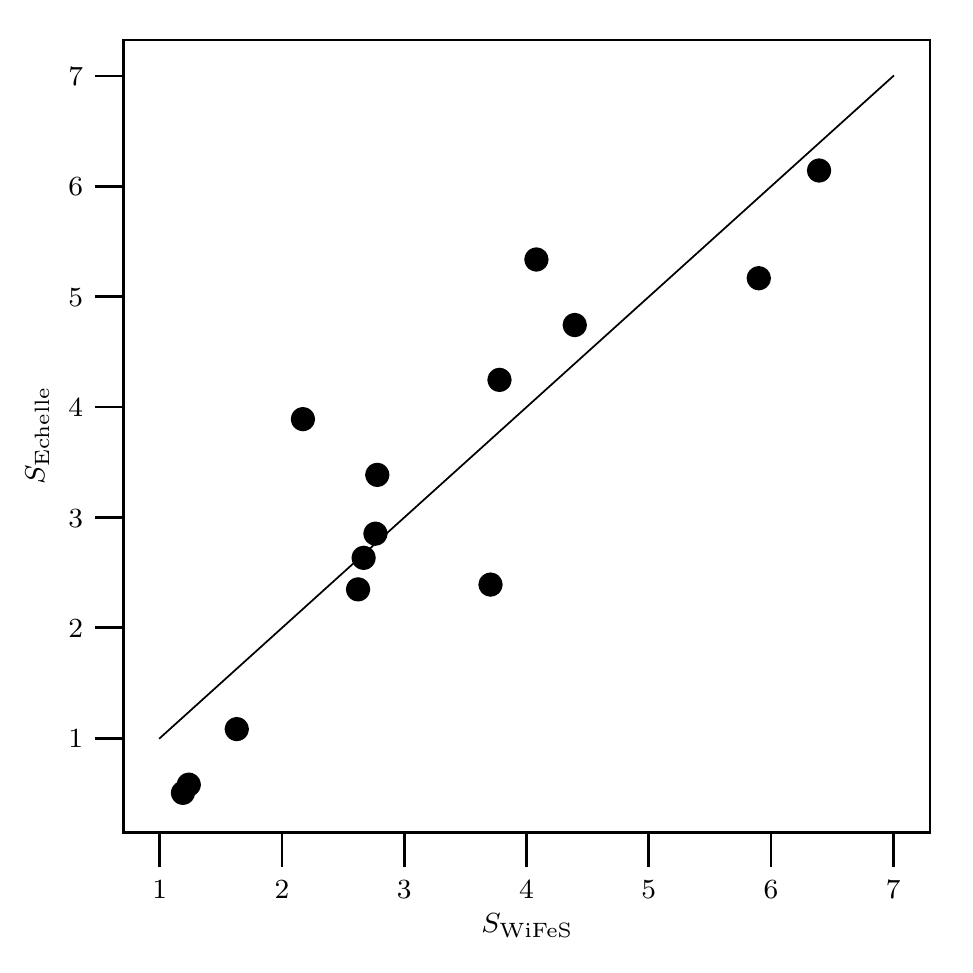}
\caption{Calibration of $\mathrm{S_{Echelle}}$ with 19 stars that were measured with both instruments.
}
\label{fig.calibrate_S_index_echelle}
\end{figure}

Note that $\mathrm{S_{Echelle}}$ and $\mathrm{S_{WiFeS}}$ are on the same scale and directly comparable. We only use a separate notation here to avoid confusion.
The relation between $\mathrm{S_{Echelle}}$ and $\mathrm{S_{WiFeS}}$ (Fig. \ref{fig.calibrate_S_index_echelle}) is suffering from a scatter for various reasons, e.g. low signal-to-noise ratio in the Echelle spectra, time variability and error propagation from the WiFeS S-index calibration.

\bsp	
\label{lastpage}
\end{document}

%% file: table.tex
4702194830625123712 & 00171443-7032021 & 11.32 &    4.31  & 1.99   & 3.95 & 20181123 & Echelle  &    1 &   14 &     12.7 &     1.5 &    -4.46 &     2.00 &     0.06 & False & False   \\
2321852448270102912 & 00252986-2834267 & 12.20 &    4.33  & 2.01    &     & 20181116 & Echelle  &      &   18 &     63.1 &     1.5 &          &     3.18 &     0.11 & False & True   \\
2375044419935775744 & 00265752-1344580 & 12.93 &    3.65  & 1.67   & 2.54 & 20190805 & WiFeS    &   19 &   62 &     -3.6 &     3.0 &    -4.72 &    -0.63 &     0.03 & False & False   \\
2375044419935775744 & 00265752-1344580 & 12.93 &    3.65  & 1.67   & 2.54 & 20190828 & WiFeS    &    5 &   26 &    -10.9 &     3.0 &    -4.60 &    -0.70 &    -0.04 & False & False   \\
2315841869173294080 & 00275023-3233060 & 11.69 &    5.42  & 2.69    &     & 20181116 & Echelle  &    1 &   15 &      7.9 &     1.5 &    -3.84 &     4.98 &     0.09 & False & True   \\
2315841869173294208 & 00275035-3233238 & 11.92 &    5.57  & 2.80    &     & 20181116 & Echelle  &    2 &   17 &      7.2 &     1.5 &          &    44.73 &     0.11 & False & True   \\
2348361020780579072 & 00311906-2334452 & 13.00 &    4.80  & 2.31   & 2.19 & 20190720 & WiFeS    &   12 &   45 &     -6.9 &     3.0 &    -5.05 &    -0.33 &     0.03 & False & False   \\
2802667100685710080 & 00362434+2142410 & 12.95 &    3.14  & 1.39   & 1.16 & 20190720 & WiFeS    &   19 &   56 &      0.4 &     3.0 &    -4.61 &    -0.50 &    -0.05 & False & False   \\
2555374905394726400 & 00365170+0535146 & 12.31 &    3.16  & 1.41   & 6.83 & 20181125 & Echelle  &    2 &   19 &     -1.5 &     1.5 &    -4.32 &    -0.79 &    -0.06 & False & False   \\
5000558443376465408 & 00393579-3816584 & 11.36 &    4.81  & 2.28   & 1.06 & 20181123 & Echelle  &    1 &   16 &      0.3 &     1.5 &    -4.17 &     3.18 &     0.07 & False & True   \\
2779822783818066304 & 00402518+1521397 & 12.80 &    3.59  & 1.64    &     & 20190828 & WiFeS    &   10 &   42 &    -36.0 &     3.0 &    -4.61 &    -0.34 &    -0.04 & False & False   \\
2345626677097190784 & 00453393-2433206 & 12.82 &    4.96  & 2.40   & 1.59 & 20190720 & WiFeS    &   14 &   49 &     -1.1 &     3.0 &    -4.69 &     0.61 &    -0.06 & False & True   \\
4925517255818631552 & 00524536-5048546 & 13.04 &    3.60  & 1.65   & 1.64 & 20190805 & WiFeS    &   15 &   51 &      7.8 &     3.0 &    -4.78 &    -0.65 &     0.01 & False & False   \\
2808754198920026112 & 00541900+2715306 & 12.10 &    4.70  & 2.28   & 3.37 & 20190710 & Echelle  &    1 &    7 &     -9.1 &     1.5 &          &    -0.03 &          & False & False   \\
2808754198920026112 & 00541900+2715306 & 12.10 &    4.70  & 2.28   & 3.37 & 20190710 & Echelle  &    1 &   12 &     -6.8 &     1.5 &    -4.76 &     0.01 &     0.35 & False & False   \\
2346249997110899840 & 00563394-2255454 & 13.17 &    4.47  & 2.15  & 15.55 & 20190720 & WiFeS    &   13 &   46 &      8.8 &     3.0 &    -4.25 &     1.16 &     0.07 & False & True   \\
2346249997110899840 & 00563394-2255454 & 13.17 &    4.47  & 2.15  & 15.55 & 20190828 & WiFeS    &    2 &    9 &     14.9 &     3.1 &    -5.02 &     1.17 &     0.14 & False & True   \\
307792101054740224  & 00584973+2752234 & 12.98 &    3.13  & 1.41  & 18.81 & 20190720 & WiFeS    &   17 &   54 &     10.6 &     3.0 &    -4.80 &    -0.82 &    -0.00 & False & False   \\
5034700237924390016 & 01084572-2551400 & 8.40 &    2.26  & 1.01    &     & 20200912 & WiFeS    &   76 &  157 &     21.5 &     3.0 &    -4.43 &    -1.09 &     0.05 & False & False   \\
5039642851928932608 & 01192734-2621549 & 12.28 &    5.19  & 2.59    &     & 20181117 & Echelle  &    0 &   14 &      5.8 &     1.5 &          &     5.97 &     0.12 & False & True   \\
5039642851928932608 & 01192734-2621549 & 12.28 &    5.19  & 2.59    &     & 20181117 & Echelle  &    1 &   15 &     10.6 &     1.5 &          &     6.38 &     0.28 & False & True   \\
4935080704877461504 & 01221098-4433502 & 12.65 &    3.81  & 1.76    &     & 20190828 & WiFeS    &   10 &   14 &      2.8 &     3.1 &    -5.59 &    -0.34 &     0.10 & False & False   \\
4916062039935185792 & 01280868-5238191 & 8.97 &    2.73  & 1.20    &     & 20200912 & WiFeS    &   65 &  150 &      7.5 &     3.0 &    -4.08 &    -0.05 &     0.19 & False & False   \\
2467225825540929920 & 01424082-0706286 & 11.26 &    3.17  & 1.44    &     & 20181117 & Echelle  &    1 &   16 &     19.3 &     1.5 &    -4.01 &    -0.67 &     0.02 & False & False   \\
291448925859674496  & 01442801+2501340 & 13.30 &    5.13  & 2.49   & 2.35 & 20190722 & WiFeS    &    9 &   35 &     -6.7 &     3.0 &    -4.32 &     2.90 &    -0.00 & False & True   \\
2575842627879030656 & 01511997+1324525 & 11.05 &    4.75  & 2.25    &     & 20181117 & Echelle  &    1 &   15 &     64.4 &     1.5 &          &     2.54 &     0.06 & False & False   \\
5135908840152003584 & 01531133-2105433 & 10.49 &    4.71  & 2.21    &     & 20181117 & Echelle  &      &   26 &     82.8 &     1.5 &          &     2.85 &     0.08 & False & True   \\
2463211714746424576 & 02001277-0840516 & 11.44 &    4.97  & 2.40    &     & 20181117 & Echelle  &    1 &   14 &      3.0 &     1.5 &    -3.74 &     5.43 &    -0.05 & False & True   \\
4967935143107585152 & 02052304-3631261 & 13.15 &    4.06  & 1.90  & 26.43 & 20190828 & WiFeS    &    4 &   23 &      7.8 &     3.0 &    -4.70 &    -0.30 &     0.02 & False & False   \\
4967935143107585152 & 02052304-3631261 & 13.15 &    4.06  & 1.90  & 26.43 & 20190711 & WiFeS    &   12 &   46 &     17.0 &     3.0 &    -4.65 &    -0.34 &     0.00 & False & False   \\
4713771622913507328 & 02224418-6022476 & 11.96 &    5.62  & 2.83    &     & 20181115 & Echelle  &      &   12 &          &     1.5 &    -5.34 &          &          & False & False   \\
5145553064660421120 & 02303485-1543248 & 12.00 &    5.09  & 2.51    &     & 20181117 & Echelle  &    1 &   17 &    -11.0 &     1.5 &    -4.24 &     3.80 &     0.13 & False & True   \\
82759763482044800   & 02370672+1707364 & 12.90 &    3.24  & 1.44   & 7.18 & 20190722 & WiFeS    &   21 &   60 &     15.8 &     3.0 &    -4.61 &    -0.65 &     0.07 & False & False   \\
4742040513540707072 & 02414730-5259306 & 11.08 &    5.00  & 2.40   & 2.13 & 20181115 & Echelle  &    0 &   13 &     11.7 &     1.5 &    -4.02 &     4.89 &     0.09 & False & True   \\
22338644598132096   & 02442137+1057411 & 10.31 &    4.34  & 1.96   & 3.07 & 20181125 & Echelle  &    2 &   27 &      4.2 &     1.5 &          &     2.11 &     0.45 & True & True   \\
22338644598132096   & 02442137+1057411 & 10.31 &    4.34  & 1.96   & 3.07 & 20190112 & WiFeS    &  218 &  335 &     15.1 &     3.2 &          &     1.79 &     0.44 & True & True   \\
114339519842832256  & 02491952+2521392 & 11.94 &    3.44  & 1.53  & 10.84 & 20181129 & Echelle  &    1 &   21 &      7.0 &     1.5 &    -4.31 &     1.05 &     0.18 & False & False   \\
5160497631000659968 & 02522075-1134484 & 13.98 &    3.36  & 1.51    &     & 20190828 & WiFeS    &    2 &   15 &     22.5 &     3.0 &    -5.06 &    -1.00 &    -0.02 & False & False   \\
5049234888291201280 & 02540274-3554166 & 7.91 &    2.59  & 1.17    &     & 20200912 & WiFeS    &   68 &  157 &     57.7 &     3.0 &    -4.51 &    -0.88 &    -0.01 & False & False   \\
4748158986511426688 & 02543316-5108313 & 11.14 &    4.77  & 2.27   & 2.88 & 20181123 & Echelle  &    1 &   15 &     11.4 &     1.5 &    -4.11 &     2.46 &     0.16 & False & True   \\
28705916434646656   & 02544314+1308519 & 13.50 &    3.75  & 1.70   & 2.97 & 20190720 & WiFeS    &   12 &   42 &     18.8 &     3.0 &    -4.49 &    -0.21 &     0.02 & False & False   \\
5078121017258751104 & 02565697-2331065 & 12.60 &    4.59  & 2.17   & 2.63 & 20181201 & Echelle  &    2 &   20 &    -35.7 &     1.5 &          &     2.32 &     0.03 & True & True   \\
5078121017258751104 & 02565697-2331065 & 12.60 &    4.59  & 2.17   & 2.63 & 20190114 & WiFeS    &   31 &   89 &    -24.1 &     3.0 &          &     1.58 &    -0.00 & True & True   \\
5078121017258751104 & 02565697-2331065 & 12.60 &    4.59  & 2.17   & 2.63 & 20190828 & WiFeS    &    7 &   33 &    -17.5 &     3.0 &          &     1.62 &    -0.01 & True & True   \\
110753570043689472  & 03082411+2345545 & 12.20 &    4.65  & 2.25  & 10.06 & 20181129 & Echelle  &    0 &   17 &     12.2 &     1.5 &    -3.71 &     1.21 &     0.14 & False & True   \\
31061035981716864   & 03094484+1513181 & 12.90 &    4.30  & 1.94   & 2.83 & 20190722 & WiFeS    &   16 &   58 &      9.7 &     3.0 &    -4.21 &     1.01 &     0.43 & False & True   \\